\documentclass[aps,superscriptaddress,showpacs,nofootinbib]{revtex4}
\usepackage{epsfig}
\input{epsf}

\begin{document}

%\baselineskip .45 in
%\vbox{
%\begin{flushright}
%AZPH-TH/94-20 \\
%VAND-TH/97-
%\end{flushright}

%\title{Large Transient states in heavy ion collision}
\title{Large transient states in quantum field theory
%\\ (Quark-Gluon Plasma: Wherefore Art Thou S-matrix Theory?)
}

\author{Arjun Berera}
\email{ab@ph.ed.ac.uk}
\affiliation{ School of Physics, University of
Edinburgh, Edinburgh, EH9 3JZ, United Kingdom}

\begin{abstract}

A study is made of the scattering of two large 
composite projectiles, such as heavy ions,
which are initially prepared in a pure quantum state.  It is shown
that the quantum field theoretic evolution equation 
for this system, under certain conditions, goes over in
form to the master equation of classical statistical mechanics.
Thus, the statistical mechanical description of heavy ion collision
is viewed as an implied outcome of the Correspondence Principle,
which states that in the limit
of large quantum number,
quantum dynamics goes over to classical dynamics.
This hypothesis is explored within the master equation transcription
with particular focus on the quark-gluon formation scenario.
\end{abstract}

\vspace{20mm}

%PACS numbers: 05.30.-d, 25.75.Nq, 25.75.-q
%11.10.Wx - finite T QFT, 12.38.Mh-QGP

\maketitle

In Press Nuclear Physics A 2007

\medskip

keywords: quantum to classical, master equation, heavy ion collision 

\bigskip

\section{Introduction}

\medskip

In the study of heavy ion collisions,
a fundamental question has been
brought to the fore about the substructure of matter beyond the
observed hadronic resonances.  Originally, Hagedorn showed that in
hadron collisions, the hypothesis of an intermediate gas like state
composed of hadronic resonances, the hadron gas, could
describe the $p_\perp$-distribution of
produced secondaries \cite{hagedorn}.
The advent of QCD and its success through the parton model, eventually
brought forth the conjecture that fundamental quarks
and gluons may be capable of existing in a plasma phase
at higher temperatures and densities to 
the hadron gas \cite{qcdptanal,qcdptlat,qgp}.
To test this idea, collisions between heavy nuclei have been a central
source of experimental input \cite{experiment},  
with at present the Relativistic Heavy Ion Collider (RHIC) operating
at $200$ GeV/nucleon \cite{experiment2}
and in the near future
A Large Ion Collider Experiment (ALICE) 
at LHC to operate at $5.5$ TeV/nucleon \cite{schukraft}.
So far experimental data seems
to rule out the hadron gas phase in very high energy heavy nuclei
collisions, and the data is also inconclusive about the quark-gluon
plasma phase \cite{experiment,experiment2}. 
Thus all parts of the theoretical picture have
not fit as yet, but there is suggestive evidence of a richer
particle-like substructure beyond the conventional hadron gas.
There are several differing ideas on the properties of this intermediate
state. However there is a unifying hypothesis, that 
heavy ion collision can produce large transient
states, which are comprised of constituents
intermediate to the final asymptotic state of hadrons,
and the evolution of such states can be described within a non-equilibrium 
statistical mechanical framework.  Moreover
the hypothesis is applied to individual scattering events and is not
simply a heuristic model for describing an ensemble average of events.

This paper develops the large transient state hypothesis of heavy
ion collision as a specific concept, independent of the particulars
of the constituents, such as quark-gluon plasma or etc...
The essential step to understand is how a pure quantum field
theory process can be interpreted within a statistical
mechanical description.  For this we identify a formal link between
conventional quantum field theoretic scattering and nonequilibrium
statistical mechanics, by determining the conditions necessary to
transcribe pure quantum evolution of a scattering process into the
fundamental evolution equation of nonequilibrium statistical mechanics,
the master equation.  This transcription has been understood for a long
time for general quantum mechanical systems \cite{vk,vh,rz}.
We will recast these ideas for scattering processes.
There is one previous effort we are aware of in \cite{te} that
has addressed the similar problem as in this paper
and we will attempt to place that work in context with
our own as we proceed.

Two clarifications are needed about the large transient state hypothesis. 
First the treatment in this paper does not pertain to the properties
of statistical density matrices.  Our considerations are
primarily focused on pure quantum states.
We are asking in what way they can be described in
thermodynamic and statistical mechanical terms.  This distinction
is important in the heavy ion problem since
it would be an unnatural expectation that asymptotic final
state properties, such as enhanced particle content
of certain species, was a result of the statistical
uncertainties in ones sample of initial states.
As a second clarification, our considerations are not related to
issues in quantum measurement theory.
All we are saying is that in the limit of large quantum number,
a particular equation with a classical statistical mechanical
interpretation will, to a good approximation,
reproduce the evolution of the system, which in rigor
is given by the quantum equations of motion.

A yet more primary aim of this paper is to examine the purpose
and formal interpretation of transients states in a scattering process.
The issue here is that within the 
standard formalism of scattering theory \cite{scat},
the treatment for heavy ions or for example electrons, is fundamentally
not much different.  In both cases the problem so posed is
given an initial states, i,  
on the in-state manifold,$|{\rm i},{\rm in} \rangle$, what is its
projection on a final state, f, on the out-state manifold
$\langle {\rm f}, {\rm out}|$.  This is a time independent formulation
of scattering. The solution of the problem requires determining the
change of basis matrix between the $|{\rm in} \rangle$ and 
$\langle {\rm out} |$ basis, thus the S-matrix,
\begin{equation}
S_{fi} \equiv \langle f, {\rm out}| i, {\rm in} \rangle.
\end{equation}
The S-matrix contains the maximum information
that can be obtained from any high energy scattering experiment done to
date, since all such experiments only measure the asymptotic state
before and after the collision. 

The standard procedure for computing the S-matrix is through quantum
field theory where two equivalent approaches are available, the
time dependent and time independent formalisms.
For scattering between simple systems such as electrons
or other fundamental particles, the time independent formalism
typically is used.  For the heavy ion problem,
due to the complexity of the problem
the S-matrix language is formally never used,
and the models used to describe the process rely
on a time evolution picture.
The underpinnings of such models
within formal scattering theory are not well understood.
The other purpose of this paper is to explore this direction.

For addressing both problems outlined above, a central framework
is developed for large transient states in heavy ion collision
relying on a three step procedure of projection, contraction and
evolution.
In the projection step, formal criteria are
specified for selecting events that emerge from
large transient states, since not all final states
in a heavy ion collision necessarily originate from such
intermediate states.  Associated with this step is
isolating such states within the scattering theory formalism,
so that they can be studied independent of other processes
occurring in the complete scattering event. 
In the contraction step, the description of the transient
state is addressed.  Complete theoretical knowledge about
the transient state is unnecessary, since the resolution
to which this state will be measured in the experiment is
grossly imperfect.   Specifically, in scattering experiments,
the transient state is indirectly measured through the
particle spectrum of a large multiparticle final state.
The measurement of only certain gross properties
of such final states are of interest, and so the transients states
from which they emerge also need only be described up
to some limited accuracy.  The contraction step is meant to
temper the theoretical description to a level adequate  
for the experimental measurement.
The combination of the projection and contraction steps
yield a well prescribed identification and
description of the transient state.
Finally the evolution step addresses the dynamical time
development of the transient state that has been modelled
by the previous two steps.
 
The paper is organized as follows. In Sect. \ref{sect2} 
the time dependent and
independent scattering formalisms  of quantum field theory are reviewed.
A general definition of transient states is given in Subsect.
\ref{sect2C} and
the time dependent formalism is then reexpressed in a form that is
convenient for treating transient states.  In Sect. \ref{sect3}
the idea of contracted description of dynamics is developed
and then applied to the heavy ion collision problem.
In Sect. \ref{sect4} the large transient state hypothesis of
heavy ion collision is presented in the context of the
ideas developed in the preceding sections.
In Sect. \ref{sect5} a master equation is presented that
describes the evolution
of two colliding heavy ions; the key significance of
this result, based on the ideas of contracted description from
earlier sections is, this master equation is describing the evolution
of a pure quantum state.
In Sect. {\ref{sect6} the phenomenology of heavy ion collision
is reviewed and then related to our formal construction from
the previous sections.
Finally Sect. \ref{sect7} is the Conclusion.
There are also two Appendices with some supporting material for
constructing the central result of this paper, the master
equation.
%this formulation is applied to QCD and the quark-gluon plasma
%hypothesis.  Special problems associated with using the formulation to a
%confining theory will be discussed primarily in section 8.

\section{Formal Scattering Theory Involving
Large Transient States}
\label{sect2}

As discussed in the Introduction, the formation of large transient states
in collision events may allow making predictions
about aspects of the asymptotic final states to
which they evolve.  In this Section a formalism is developed
for treating large transient states in scattering.
Such a formalism must encompass three details.
First, since transient states inherently are spacetime
concepts, they must be described within the time-dependent 
formalism of scattering.  Second, the outcome of any scattering process
in principle can be described in terms of the time independent
formalism of scattering, since 
information is needed only about the asymptotic initial and
final states and the S-matrix supplies the relevant
transition amplitude.  As such there must exist an interpretation
of large transient state formation during scattering,
in terms of the time-independent formalism.  Third,
only for some subclass of all scattering events,
the transient state description may be relevant 
and so a projection operator is needed to separate out
such events.

To develop the above three points, the outline of this section is
as follows.  The time-dependent scattering formalism first will
be reviewed within the context of the collision of two incoming
heavy ion wavepackets.  Then this approach will be related
to the time-independent scattering formalism.   With these
basic points in place, focus will turn to large transient
states.  A projection operator will be defined to formally
separate final states that evolved from large intermediate
transient states.  With this, the transient state
formally now can be isolated. At this point, it is a simple matter
to apply the time evolution operator to evolve the transient
state.  Furthermore, the projection step along with
invoking basic assumptions about large transient states allows
deducing some properties of the S-matrix elements for events with
large transient states.

\subsection{Time Dependent Scattering Theory}

Let 
\begin{equation}
| \phi_{\bf k}^a \rangle \equiv c_a^{\dagger}({\bf k}) | 0 \rangle
\label{eista}
\end{equation}
denote the single particle eigenstate of the full Hamiltonian with
three momentum ${\bf k}$, for species a, and with energy
$E_a({\bf k}) = \langle \phi_{\bf k}^a | {\hat H} | \phi_{\bf k}^a \rangle$,
where ${\hat H}$ is the full Hamiltonian of the theory.
The left hand side of
Eq. (\ref{eista}) defines the operator $c_a^{\dagger}({\bf k})$ on the
right hand side and $|0\rangle$ is the true vacuum of the theory.
Let the initial state 
of the two incoming relativistic heavy ions ($a=h_j, j=1,2$)
at $t_i \ll 0$
be two wavepackets
centered at coordinates ${\bf x}_1 = (0,0,z_1)$, 
${\bf x}_2 = (0,0,-z_2)$ with momenta 
${\bf k}_1 =(0,0,k_1)$, ${\bf k}_2 = (0,0,k_2)$ respectively. The
wavepackets are initially far apart, so that
$| {\bf x}_1 -{\bf x}_2 | \gg \Delta x_1+\Delta x_2$, where
$\Delta x_j$ is the spread in wavepacket j.  The wavepackets
are prepared such that their classical trajectories collide at $t=0$, so
that $z_1=z_2 \equiv z_i$ and $ct_i \equiv -z_i$.  This initial state evolves
in the Schrodinger  representation as
\begin{equation}
|I(t) \rangle = \exp[-i{\rm H}(t-t_i)] |I(t_i) \rangle .
\end{equation}
For $|I(t_i) \rangle$ to represent a incoming state of two particles means
that for $t_i \rightarrow -\infty$
\begin{eqnarray}
\langle I(t_i) | O({\bf x}) |I(t_i) \rangle 
\stackrel{\longrightarrow}{{t_i \rightarrow -\infty}}
\langle I_1(t_i) | O({\bf x})
|I_1(t_i) \rangle + 
\langle I_2(t_i) | O({\bf x})
|I_2(t) \rangle ,
\end{eqnarray}
where ${\rm O}({\bf x})$ is a local operator representing any observable and 
\begin{equation}
|I_j(t_i) \rangle = 
\int d^3{\bf k} f^{{\bf x_j},{\bf k_j}}({\bf k}) 
c_{h_j}^{\dagger}({\bf k})|0\rangle {\rm e}^{-{\rm i E_{h_j}}({\bf k}) t_i} ,
\end{equation}
where $f^{{\bf x_j},{\bf k_j}}({\bf k})$ is a smearing function,
which in coordinate space is peaked about $x_j$ and
in momentum space is peaked about $k_j$.
Thus for appropriate choice of smearing functions that localize the two
incoming particles far apart, as specified above, the initial state
is\footnote{For notational convenience, 
we will assume that both incoming particles and all outgoing particles
are distinguishable.  This could be extended to treat the statistics of
indistinguishable particles.}
\begin{equation}
|I^{\rm SE}_{h_1({\bf x}_1,{\bf k}_1),h_2({\bf x}_2,{\bf k}_2)}
(t_i) \rangle =
\int d^3{\bf k} d^3{\bf k}'
f^{{\bf x_1},{\bf k_1}}({\bf k})
f^{{\bf x_2},{\bf k_2}}({\bf k}')
c_{h_1}^{\dagger}({\bf k})
c_{h_2}^{\dagger}({\bf k}')|0\rangle
{\rm e}^{-i[E_{h_1}({\bf k})+E_{h_2}({\bf k}')] t_i} .
\label{inoe}
\end{equation}
The superscript ${\rm SE}$ denotes the basis of product ``single-particle
eigenstates''. In an interacting theory, such product states are not
eigenstates. However if the theory describes scattering, then such states
must be approximate eigenstates when all the particles are far apart. This
statement will be further clarified in the next subsection.
 
Let the final state after the collision at $t_f \gg 0$ be some
n-particle state of species $a_1 \ldots a_n$ with
outgoing wavepackets that are widely spaced.  The
details for representing the final state are the same as the initial
state, and we obtain
\begin{equation}
\langle F^{\rm SE}_{\{{\bf x}\},\{{\bf q}\}}(t_f) | = 
\int\prod_{i=1}^{n} d^3{\bf q}'_i
\langle 0| c_{a_1}({\bf q}'_1) \ldots
c_{a_n}({\bf q}'_n)
f^{{\bf x_1},{\bf q_1}}({\bf q}'_1) \ldots
f^{{\bf x_n},{\bf q_n}}({\bf q}'_n)
{\rm e}^{i[\sum_{j=1}^n E_{a_j}({\bf q}'_j)] t_f} ,
\label{outoe}
\end{equation}
with $\{ {\bf q}\} \equiv \{{\bf q}_1 \ldots {\bf q}_n \}$.
The amplitude $A_{fi}$ to go from the initial state at time $t_i$ to the
final state Eq. (\ref{outoe}) at time $t_f$ is
\begin{equation}
A_{fi} = \langle F^{\rm SE}_{\{{\bf x},{\bf q}\}}(t_f) |
\exp[-i {\rm H}(t_f-t_i)] 
|I^{\rm SE}_{({\bf x}_1,{\bf k}_1),({\bf x}_2,{\bf k}_2)} (t_i) \rangle_
{t_i \rightarrow -\infty, t_f \rightarrow \infty}.
\label{scatamp}
\end{equation}

\subsection{Time Independent Scattering Theory}

The time independent formulation of scattering is obtained from the time
dependent one by first reexpressing $|I^{SE}(t_i) \rangle$ and
$\langle F^{SE}(t_f) |$  in an eigenstate basis.  There are two
eigenstate bases that are particularly convenient, the in and out
basis.  They both satisfy the Schrodinger equation
\begin{equation}
{\rm H} \psi ^{{\rm in} ({\rm out})}(\{{\bf k}\})
 = {\rm E}_{\psi}({\{\bf k \}}) \psi ^{{\rm in} ({\rm out})}(\{{\bf k}\})
\label{ioham}
\end{equation}
but have different boundary conditions. The boundary conditions for the
in (out) states are such that they describe incoming (outgoing)
particles at large negative (positive) time.  The
incoming states are defined as follows.  Consider the case of two
incoming heavy ions $h_1,h_2$, with 
$\psi^{\rm in}_{h_1h_2}({\bf k}_1, {\bf k}_2)$ their
eigenstate in the in-basis.  Let
\begin{equation}
|I^{{\rm IE}}(t_i) \rangle  = 
\int d^3{\bf k} d^3{\bf k}'
f^{{\bf x_1},{\bf k_1}}({\bf k})
f^{{\bf x_2},{\bf k_2}}({\bf k}')
\psi^{\rm in}_{h_1,h_2}({\bf k}_1,{\bf k}_2)
{\rm e}^{-i(E_{h_1}({\bf k})+E_{h_2}({\bf k}')) t_i} ,
\end{equation}
where the superscript ${\rm IE}$ denotes that the
expansion basis is the ``in-eigenstates'' and the smearing functions
$f^{{\bf x},{\bf k}}({\bf k}')$ are the same as in 
eq.(\ref{inoe}).  The statement that $\{\psi^{\rm in}\}$ are in-eigenstates
means, in addition to Eq. (\ref{ioham}),
\begin{equation}
\langle \xi |I^{\rm IE}_{{\bf k}_1,{\bf k}_2}(t_i) \rangle
\rightarrow_{t_i\rightarrow -\infty}
\langle \xi |I^{\rm SE}_{{\bf k}_1,{\bf k}_2}(t_i) \rangle ,
\label{incond}
\end{equation}
where $| \xi \rangle$ is any arbitrary fixed state in the Hilbert space.
Similarly for out-eigenstates, let
\begin{equation}
\langle F^{{\rm OE}}_{\{{\bf q}\}}(t_f) |  =
\prod_{i=1}^{n} \int d^3{\bf q}'_i
f^{{\bf x_1},{\bf q_1}}({\bf q}'_1) \ldots
f^{{\bf x_n},{\bf q_n}}({\bf q}'_n)
\psi^{\rm out}_{a_1, \ldots, a_n}({\bf q}'_1, \ldots, {\bf q}'_n)
{\rm e}^{i(\sum_{j=1}^n E_{a_j}({\bf q}'_j)) t_f} ,
\end{equation}
where OE denotes the basis of ``out-eigenstates''. In analogy with
Eq. (\ref{incond}), in this case the condition is
\begin{equation}
\langle F^{\rm OE}_{\{{\bf q}\}}(t_f) | \xi 
\rangle_{t_f \rightarrow \infty}
\longrightarrow
\langle F^{\rm SE}_{\{{\bf q}\}}(t_f) | \xi \rangle .
\label{outcond}
\end{equation}

The scattering amplitude Eq. (\ref{scatamp}) in terms of the in-out
basis becomes
\begin{equation}
A_{fi} = \int \prod_{j=1}^{n} d{\bf q}'_j
f^{{\bf x_j}{\bf q_j}}({\bf q}_j')
\int d{\bf k}'_1 d{\bf k}'_2 
f^{{\bf x_1}{\bf k_1}}({\bf k}_1')
f^{{\bf x_j}{\bf k_2}}({\bf k}_2')
S_{\{{\bf q},a\},\{{\bf k},h\}} ,
\end{equation}
where 
\begin{equation}
S_{\{{\bf q},a\},\{{\bf k},h\}}=
\langle \psi^{\rm out}_{\{a\}}(\{{\bf q}\}) |
\psi^{\rm in}_{\{h\}}(\{{\bf k}\})
\label{smatrix}
\end{equation}
is the S-matrix.  This completes the demonstration of relating
the time dependent and independent scattering formalisms in quantum
field theory.

\subsection{Transient State Description of Scattering}
\label{sect2C}

In this Subsection the time-dependent scattering formalism is reexpressed
to allow a convenient description of states that are intermediate
to the asymptotic incoming and outgoing states,
which hereafter will be referred to as transient states. 
The construction below is
purely kinematic, thus is nonspecific to the nature of the transient
states. However, to help fix ideas, this construction will be developed in the
context of the heavy ion collision scenario.

\subsubsection{Definition - Projection}
\label{proj}

According to quantum mechanics, measuring a system in a state 
$\langle F |$ at time $t_F$ results in the filtering or reduction of the
wavefunction $\Psi (t_F)$ at time $t_F$.  Quantum dynamics implies that
an appropriate filter or projection operator 
$P_F^t$ can be applied to $\Psi(t)$ at any time 
$t< t_F$ such that the projected state at t,
$\psi_{P_f}(t)$, will evolve into $|F\rangle$
at $t_F$, with predicted amplitude 
$\langle F | \psi(t_F) \rangle$.  
Specifically let
\begin{equation}
P^t_F \equiv  e^{i{\rm H}(t_F-t)} |F \rangle
\langle F| e^{-i{\rm H}(t_F-t)},
\end{equation}
then
\begin{eqnarray}
\langle F| e^{-i{\rm H}(t_F-t)} | P_F^t \Psi(t) \rangle 
& \equiv & \langle F| e^{-i{\rm H}(t_F-t)} | \Psi_{P_F}(t) \rangle \\
& = & \langle F| \Psi(t_F) \rangle
\label{f1}
\end{eqnarray}
and
\begin{equation}
\langle F'| e^{-i{\rm H}(t_F-t)} | P_F^t \Psi(t) \rangle =0 ,
\label{f2}
\end{equation}
if $\langle F' | F \rangle =0$.

A final state $\langle F|$ at time $t_F$ will be said to have been
composed of N-transient states at time $t<t_F$ with respect to
a explicit basis $\{\xi \}$ if $\psi_{P_f}(t)$ factorizes
into an N-product state of the form  
\begin{equation}
\Psi_{P_F}(t) = \prod_{j=1}^N \Psi^{\xi}_{T_{F_j}}(t).
\label{trandef}
\end{equation}
The time interval $\Delta t_{P_{F_j}}$ over which a state
$\Psi^{\xi}_{T_{F_j}}$ retains a factorized form from the rest of the
state vector defines its lifetime.

%This definition of transient states is consistent with the quantum
%mechanical reduction postulate, since any state can be considered as
%transient until it is measured.  
Note that the decomposition of
transient state Eq. (\ref{trandef}) is dependent on the choice of basis
$\{\xi\}$. 
Also note that by this definition, the state 
$\langle F|$ is not restricted to the asymptotic manifold of final
states.  $\langle F|$ may characterize an intermediate configuration,
thus permitting identification of different stages in the evolution of a
transient state.

Aside from the lifetime $\Delta t $, transient states are described with
the standard set of operator observables, energy, momentum internal
charge and so on.  By large transient state, we always mean 
a large many-body
transient state.  To identify a transient state as large requires the
existence of a number density operator ${\hat N}_{\xi'}({\bf x})$
that acts on some subspace $\{\xi'\}$ of
the $\{\xi\}$-basis.  A transient state $\Psi_T$ is large if the number
density of $\xi'$-quanta is sufficiently large within a sufficiently
large volume V,
\begin{equation}
\langle \Psi_T| {\hat N}_{\xi'}({\bf x}) | \Psi_T \rangle - 
\langle 0 | {\hat N}_{\xi'}({\bf x}) | 0 \rangle > 
n^{{\rm min}}_T \ \ \ \ {\rm for} \ \ \ \ {\bf x} \in  V > V^{\rm min}_T,
\label{pnumber}
\end{equation}
where $|0\rangle$ is the vacuum state.
The specific values of $n^{\rm min}_T$ and $V^{\rm min}_T$ depend on the
specific dynamical theory.

\subsubsection{Description}
\label{projdes}

A description of transient states requires dividing the collision process
into a pre, during and post collision time period.  In the pre-collision
period, the two incoming wavepackets are mutually non-interacting, thus
are well described by $|I^{\rm SE}(t) \rangle$ up to the time
$t=t_c^- < 0$. Here $t_c^-$ is defined as the time when the rms spread of
both incoming wavepackets first make contact.  The post-collision period
begins as $t_c^+>0$ when the rms spread of the two wavepackets would no
longer overlap had the two wavepackets never interacted.  If both wavepackets
are symmetrical about their respective centers then $t_c^+=|t_c^-|$.
The transient states are formed during the collision time period
$t_c^- < t < t_c^+$ and evolve during the post-collision period as
isolated localized systems i.e. blobs.  The formation can be of any number
of isolated transient states. Generically, in heavy ion collision three
transient states are hypothesized, one in the central region 
($\sim 3$ units of central rapidity) and two
"fireballs" in the right and left fragmentation 
regions respectively \cite{akm}.
During the post-collision period for each transient state j, there is a
hadronization time $t_{h_j} > t_c^+$. For $t> t_{h_j}$ the state j is
simply described in terms of the asymptotic single-particle eigenstate basis 
$|F^{SE}_j(t>t_{h_j}) \rangle$. 

\subsubsection{Formalism}

In terms of time-dependent scattering theory, the transient state
description is expressed as follows.  For $t<t_c^-$ as said above
the state is $|I^{\rm SE}(t) \rangle$.  The transient state
begins, by definition, at $t_c^-$ starting with the first stage, the
formation state. During the formation
period, $t_c^+>t>t_c^-$, the state vector evolves as
\begin{equation}
\psi_T(t) \equiv 
{\rm e}^{-i {\rm H}(t-t_c^-)} | I^{\rm SE}(t_c^-) \rangle.
\end{equation}
Hard interactions occur predominately in the formation period.
Anytime after $t_c^-$ the formation state in general
may separate into multiple
transient states.  Although no separation will be sharp, it
is convenient to express approximately separate states as such.
The phenomenological nature of the
transient state approach reflects in the lack of a prescriptive
procedure for
implementing the separations.  
Hadron phenomenology indicates that collisions typically separate
spatially into three regions, central (C) and a left (L) and right (R)
fragmentation region.  For a final state $\langle F^{\rm SE} |$ that
evolved from an intermediate large transient state, during the formation
period there should be a single transient state.  At the end of the
formation period, $t_c^+$, this transient state should spatially separate
approximately into these three components.   In any basis $\{\xi\}$ in which the
elementary excitations describing the transient state
are local, the separated state can be expressed in
the factorized form 
\begin{equation}
{\rm e}^{-i {\rm H}(t_c^+-t_c^-)} | I^{\rm SE}(t_c^-) \rangle
= | \psi_C^{\{\xi\}I}(t_c^+)
\psi_L^{\{\xi\}I}(t_c^+)
\psi_R^{\{\xi\}I}(t_c^+) \rangle ,
\label{3trans}
\end{equation}
where the supercript I denotes the incoming asymptotic state from
which this transient state evolved.  QCD effects will
smear this sharp separation, but
an approximate
separation is justified if the transient states are
sufficiently large.
Assuming this approximation,
for $t>t_c^+$ the evolution of each transient state in 
Eq. (\ref{3trans}) is independent
\begin{equation}
|\psi_J^{\{\xi\}I} (t>t_c^+) \rangle = 
{\rm e}^{-i {\rm H}(t-t_c^+)} | \psi_J^{\{\xi\}I}(t_c^+) \rangle ,
\end{equation}
where $J = C,L,R$.  Define
\begin{equation}
A_{\xi,\psi_J^I}(t-t_c^+) =
\langle \xi | 
{\rm e}^{-i {\rm H}(t-t_c^+)} | \psi_J^I(t_c^+) \rangle
\end{equation}
as the probability amplitude for the transient state $J$ at $t_c^+$ to be in
state $|\xi \rangle$ 
at $t$.  At the end of the transient state period
$t_{h_j}$, each of the three spatially separated
transient state will hadronize independently.
The final asymptotic state can be factorized into its
C,R and L regions as 
$\langle F^{SE}|  = \langle F_L F_C F_R|$, with
amplitude to go into a particular 
final state $\langle F_J^{SE}(t_{h_j})|$,
\begin{equation}
A_{F_J^{\rm SE},\psi_J^{\{\xi\}I}} \equiv
A_{F_J^{\rm SE},\psi_J^{\{\xi\}I}}(t-t_c^+)|_{t-t_c^+ \rightarrow \infty} =
\langle F_J^{\rm SE}(t)| 
{\rm e}^{-i {\rm H}(t-t_c^+)} | \psi_J^{\{\xi\}I}(t_c^+) 
\rangle_{t \rightarrow \infty} .
\end{equation}
For scattering process that are well described by a transient
state time-dependent approach in a basis $\{\xi\}$,
there corresponds in the
time-independent approach a factorized form for the respective
S-matrix elements in the same
basis, \footnote{It is worthwhile to compare the transient state 
description above to the one
in hard scattering \cite{css}.
For hard scattering, the parton model hypothesis implies
a separation at
$t_c^-$ as
a two component transient state
\begin{equation}
|I(t_c^-) \rangle = |H\rangle \times |J\rangle,
\label{hardtr}
\end{equation}
where $|H\rangle$ is the partonic state that participates in the hard
interaction and $|J\rangle$ is the state of the two beam jets.  
For this case, pQCD can be used to determine the validity 
of the separation.
It is well know from pQCD that an initial state of the form
eq (\ref{hardtr}) rapidly loses its factorized form due to soft and
collinear gluon exchange between $|H\rangle$ and $|J \rangle$,
which leads to large infrared corrections.. However
detailed calculation from pQCD shows that these divergences cancel for
the probability amplitude squared once it is summed over a specific class
of final states.  This expectation of pQCD has been convincingly verified
by experiment in particular for inclusive hard scattering.
Thus the transient state ansatz 
Eq. (\ref{hardtr}) is good for extracting inclusive information about hard
scattering.  In contrast, as will be seen, the transient state hypothesis
of heavy ion collision addresses individual events.}

\begin{equation}
S_{FI} = \prod_{J=C,L,R} A_{F_J,\psi_J^{\{\xi\}I}}.
\end{equation}

\section{Contracted Quantum Dynamics}
\label{sect3}

Once the large transient state has been isolated 
with the projection operator of Subsect. \ref{proj},
focus turns to the description of the state itself.
Recall a motivation for considering events with large transient states
is that due to their large size, their dynamics may have simplifying features
and at the same time such a state could dominate many features
of the final asymptotic states.
A central point here is that complete detail about the
large transient state is not necessary, but rather
a contracted description suffices.  In this Section,
such a contracted description for large transient states
is obtained.  This will start first with a elementary discussion
of contracted quantum dynamics, with
specific emphasis on the implication of the large
number limit.

\subsection{Definition - Contraction}
\label{contract}

Almost any measurement in physics results in a contracted description of
that system.  This is both because not all observables of the system
are measured and because of the ones that are, the measurement
uncertainties exceeds the intrinsic quantum mechanical uncertainties.
To a given measurement, all theoretical descriptions of the system in
question are equally fundamental if they can correctly predict the
values of the specific measured observables within their measured
uncertainties.  The most fundamental theory is the one that correctly
predicts all observables of any system and within the intrinsic
uncertainties set by quantum mechanics.  Less fundamental theories are
derived consequences of this most fundamental theory in which some
information has been removed.  These less fundamental theories are
called contracted descriptions. They form a hierarchy in which one
contracted description is derived from a higher one and so on up to the
most fundamental theory.

In quantum mechanics a physical process is generally
analyzed by the following sequence of steps.
The measurement is described through a set of operator
observables $\{{\hat O}_j\}$, which in general are measured at
various times $t_l$, from which the experimental data
emerges as expectation values $\{O_j(t_l)\}$ with measurement
uncertainties $\{\Delta O_j(t_l)\}$.  The objective then is to explain
this experimental data from theory.  In quantum mechanics,
this implies determining the wavefunction $\psi(t)$ of the system.
Any wavefunction solution $\psi$ would be considered adequate, if all
$\{O_j^{th}(t_l)\}$ agree with $\{O_j^{exp}(t_l')\}$, where
\begin{equation}
O_j^{th}(t_l) = \langle \psi(t_l) |{\hat O_j}|\psi(t_l)\rangle ,
\end{equation}
and within the measurement uncertainties $\{\Delta O_j(t_l)\}$, 
so
\begin{equation}
|O^{th}_j(t_l) - O^{exp}_j(t_l)| \leq \Delta O_j(t_l) .
%[\langle \psi | {\hat O}_j^2 | \psi \rangle -
%\langle \psi | {\hat O}_j | \psi \rangle^2 ]^{1/2} \leq \Delta O_j
\label{deloi}
\end{equation}

The most fundamental description would require determining
all the energy eigenstates $\{\Psi_n\}$ of the system from which the 
system wavefunction in general would be expressed as
\begin{equation}
\psi(t) = \sum_n a_n \phi (E_n) \exp(-iE_nt) ,
\label{psit}
\end{equation}
with the expansion coefficients $\{a_n\}$ fixed from experimental data.
If measurements at some initial time $t_{0}$ are done
of a complete set of operator observables and up to minimum
quantum mechanical uncertainty, then all expansion
coefficients $\{a_n\}$ could be precisely determined.
On the other hand, if the measurement is less complete,
then the state can not be precisely specified.
In this case the wavefunction $\psi(t)$ only can be approximately
known, which means either the expansion coefficients
can be determined only up to some uncertainty $\{\Delta a_n\}$,
or the states of the expansion basis in Eq. (\ref{psit}) 
are only approximate
energy eigenstates, resolved up to measurement uncertainty
$\Delta U$, or a combination of both.

In this Section an explicit contracted description will be obtained
for the case of large transient states in heavy ion collisions.
For this, the notion of ``similar state'' will be useful.  Consider
a wavefunction expanded at some time $t_i$
in an arbitrary basis $\{ |\alpha \rangle \}$ as
\begin{equation}
\psi =\int \sum_j c_j | \alpha_j \rangle .
\label{psieca}
\end{equation}
Two states $|\alpha_1\rangle$, $|\alpha_2 \rangle$ are 
defined to be similar at time $t_i$ with
respect to a specified set of observables $\{ \hat O\}$ if conditions 
Eqs. (\ref{deloi}) remain valid at this time
$t_i$  after the replacement
$|\alpha_1 \rangle \rightarrow |\alpha_2 \rangle$ or
$|\alpha_2 \rangle \rightarrow |\alpha_1 \rangle$ in Eq. (\ref{psieca}).
Thus the specified set of operators and measurement uncertainties
partition the Hibert space into cells of similar states at each
time $t_i$.  The degree of partitioning depends on the choice of
Hibert space basis.  A poor choice results in very few similar states
per cell, whereas a good choice results in a large number of
similar states per cell.

\subsection{Relevance to Heavy Ion Collision}

The formation of large transient states in a heavy ion
collision requires a contracted description both for theoretical
and experimental reasons.  In regards the former, the unknown
confinement mechanism relinquishes ability to
compute many details about the transient state.  Even without
compounding this problem, the complexity of such large states
only can be understood, in practice, by approximation methods.
Here the first implementation of contraction arises through use
of the quasiparticle concept.  This concept plays a central
role since a tacit assumption is that the main
properties of the large
transient state are controlled by a large density of nearly free 
particle-like states.  The quark-gluon plasma picture of
the large transient state is a common example of application of the
quasiparticle concept.  However as discussed in sections to come,
a more general description underlies this.

From the perspective of experiment, a contracted description also
is required, since a large transient state leads to a large
multiparticle final state.  Although in principle such states can
be measured with precise resolution, limited only
by quantum mechanical uncertainties, in practice this is out of
the question.  Moreover, extracting information about only
the transient state from the complete final state data,
intertwines experimental measurement with theory.  
Ultimately a middleground is sought, where information from 
experiment and theory are in mutual balance.  The large transient
hypothesis is testable since it falls within this middleground.
Its underlying premise is that the large number limit dominates
the dynamics, in which case theoretical methods are
available to treat it and at the same time the size of the state
should leave a identifiable signature in the final state
data.  

Thus observables measured in the contracted description
minimally will include total energy U \footnote{In this paper
energy is denoted in a few different ways, depending in what context
it is used.  The quantum mechanical Hamiltonian is denoted as
usual as ${\hat H}$ and the n-th energy eigenstate is denoted
as $E_n$.  On the other hand, energy expectation values, such
as those giving total energy or energy within some specified
region etc... will be denoted as $U$;  this choice follows a
convention for energy often used in thermodynamics and statistical
mechanics \cite{fermi2,kittel}, and so carries the right suggestive meaning also for
this paper.}
and particle number,
which expressed in thermodynamic language can be termed
entropy S.  The volume of the transient state also
could be measured \cite{hbt}, from which combining with U,
a temperature T can be assigned.  The measurements
of energy and entropy also may be further resolved into bins
of different momentum,
particle species, etc... of the respective particles.

\subsubsection{Role of time-energy uncertainty in measurement}
\label{timeup}

The concept of a transient state implies temporal evolution
with observations of the state at successive intervals of times.
However in a scattering experiment, observations only
occur twice, at asymptotic times before, 
$t_i$, and after, $t_f$, the collision, so that
transient state formation during a collision process
is a notion that can be tested only indirectly.  The
putative transient state must be modelled, with
the predictions that each model yields
only at asymptotic large time testable with experiment.
This clearly leaves considerable room for degeneracy amongst
candidate models, and so speaks to a central difficulty
of the transient state problem of heavy ion collision.
Nevertheless, within the mathematical construct, the transient state
takes on a dynamical reality, and all models that correctly
predict the outcome of the collision experiment are viable dynamical
scenarios during the intermediate time period.

The first step in determining models for the transient state
is specifying the maximum energy resolution to which the actual state
is measured, since any model need only be valid up
to this resolution.
If the lifetime of the transient state is $\Delta t_{tr}$,
then the uncertainty principle dictates that the minimum
energy uncertainty to which the state can be
determined is $\delta U \sim 1/\Delta t_{tr}$.
If a measurement of the state were made to such a refined
resolution, then comparison to theory would require knowing the
complete quantum mechanical wavefunction of the state.
In a heavy ion collision experiment, even if in principle
the final state were measured with perfect resolution,
it still would not imply the same holds for the transient state.
This is because the transient state can not be measured
in isolation, since it is surrounded by many
other processes that comprise the entire collision event.
The final state that is observed in the detectors
contains a vast mixture of information about all these processes.
Separating out the portions due to the transient state would 
introduce a measurement uncertainty in the energy $\Delta U$ much larger
than the intrinsic uncertainty, $\Delta U \gg \delta U$.
The strength of the large transient
state hypothesis is that for energy resolution
$\Delta U \gg \delta U$, but still $U > \Delta U$,
the large number limit dictates the properties of the
transient state, so that neither perfect experimental nor
theoretical knowledge about the collision process is
necessary.
Thus models
of putative transient states must be accurate only to within
the measurement uncertainty $\Delta U$, with dynamical evolution 
in the model treated only in time steps of
size $\Delta t \stackrel{>}{\sim} {\hbar}/\Delta U$.

To quantify the above inequalities, consider the 
evolution of states formed in the central
region in a heavy ion collision.  It is believed that
for events in which large transient states formation
is possible in the central
region \cite{bj,mcl}, 
the total energy of such states is $U \approx (5-1000) {\rm GeV}$
and $\Delta t_{tr} \approx (10-1000) {\rm fm/c}$.
Since in a collision experiment information about the transient state
must be inferred from the final state particle spectrum,
the minimum measurement uncertainty of the
transient state energy can be set as that of a typical minijet 
$\Delta U \stackrel{>}{\sim} 1 {\rm GeV}$.  For these values, it implies
$\delta U \approx h/\Delta t_{tr} \approx (10^{-2} - 10^{-4}) {\rm GeV} \ll \Delta U \approx 1 {\rm GeV}$ 
and the evolution  time step must be
$\Delta t  \stackrel{>}{\sim} 0.2 {\rm fm/c}$.
%> (10^{-2} - 10^{-4}) {\rm fm/c}$.

\subsection{Contracted description of large transient states}
\label{contdy}

This subsection formulates the contracted description
of large transient states with specific reference
to heavy ion collision.  Consider a set of operator
observables $\{{\hat O}_j\}$ that are experimentally measured in
the final state, with measurement uncertainties $\{\Delta O_j \}$.
Amongst the observables would
be the total energy $U$, entropy $S$, perhaps also local
versions of these operators in smaller volume regions
of the putative transient state, and other observables.
All the observables are macroscopic in that the
intrinsic quantum mechanical uncertainties
\begin{equation}
\delta O_j \equiv
[\langle \psi | {\hat O}_j^2 | \psi \rangle -
\langle \psi | {\hat O}_j | \psi \rangle^2 ]^{1/2},
\label{qmuncert}
\end{equation}
are smaller than the measurement uncertainties $\Delta O_j$,
\begin{equation}
\Delta O_j \Delta O_l \gg \delta O_j \delta O_l,
\label{dodo}
\end{equation}
for all $j,l$.
Within our mathematical construct of the transient state, we wish to
model evolution in time steps of 
$\Delta t \stackrel{>}{\sim} {\hbar}/\Delta U$ with 
operator expectation values resolved at each time step
up to the measurement uncertainties $\{\Delta O_j \}$ .

\subsubsection{Slowly varying operators}

Recalling the uncertainty principle relation
\begin{equation}
\delta U \delta O_j \stackrel{>}{\sim} \langle \psi| 
[{\hat H},{\hat O}_j] 
|\psi \rangle \approx  \langle \psi| {\dot {\hat O}}_j | \psi \rangle,
\label{commr}
\end{equation}
Eq. (\ref{dodo}) implies
\begin{equation}
\Delta U \Delta O_j \gg \langle \psi | {\dot {\hat O}}_j |\psi \rangle ,
\label{dudo}
\end{equation}
where $\psi$ is any arbitrary states.
Also, the observation time interval $\Delta t$ is dictated
through the uncertainty principle relation
\begin{equation}
\Delta U \Delta t \stackrel{>}{\sim} {\hbar}.
\end{equation}
Thus for measurement time intervals larger than the uncertainty
principle lower bound, but not in excess to Eq. (\ref{dudo})
it implies
\begin{equation}
\Delta U \Delta O_j \sim \Delta O_j \hbar/{\Delta t},
\end{equation}
so that from Eq. (\ref{dudo})
\begin{equation}
\Delta O_j \gg \langle \psi | {\dot O}_j | \psi \rangle \Delta t/\hbar ,
\label{slowv}
\end{equation}
for any state $\psi$.
Operators satisfying Eq. (\ref{slowv})
are called ``slowly varying'' in \cite{vk}.
In terms of Eq. (\ref{psieca}) if all operators used in partitioning similar
states are slowly varying, then it implies this partitioning remains
valid over several measurement time intervals.
For a  partitioning of states based on such a set of
slowly varying operator observables, in Sect. \ref{sect5} it
will be shown that the state vector obeys 
a evolution equation that is classical in form.
There are several details related to the above construction
requiring comment and they will be addressed in the three
subsections that follow.  

\subsubsection{Approximate diagonalization}

Amongst the operator observables, the energy operator has a special
role, since energy will be conserved.   The measurement of the system's
energy at value $U_N$ with uncertainty
$\Delta U$ separates the energy eigenstates into shells N.
If energy is the only measured operator
then each shell $N$ has a set of approximate energy eigenstates 
$\{\psi_{Nn}\}$ such that
\begin{equation}
\langle \psi_{Nn} | {\hat H} | \psi_{N'n'} \rangle =
[U_N + O(\Delta U)] \delta_{NN'}\delta_{nn'} + O(\ll \Delta U).
\end{equation}
Here the magnitude of the off-diagonal elements $O(\ll \Delta U)$
are determined by the conditions Eqs. (\ref{qmuncert}) and (\ref{dodo}) 
which implies
for any $m$, $\sum_{n, n\ne m} |H_{mn}|^2 \equiv \delta U^2 \ll \Delta U^2$.

Now consider another operator ${\hat O_1}$.  Due to relations
Eqs. (\ref{dodo}) and (\ref{commr}), up to measurement 
uncertainty, ${\hat H}$ and
${\hat O_1}$ commute and so can be simultaneously
approximately diagonalized as follows.
A diagonalization in general would require all the approximate eigenstates
$\{\Psi_{Nn}\}$ from all the shells N, but first
it will be shown
that the diagonalization done here can be achieved in each shell N
with only the approximate eigenvectors 
in that particular shell.  From Eq. (\ref{commr})
choosing $\psi = a \Psi_{Nn} + b \Psi_{N'm}$ where for
normalization $|a|^2 + |b|^2 = 1$,
consider a particular matrix element
\begin{equation}
({H}{O}_1 - {O}_1 {H})_{N'm,Nn} \approx (E_{N'm}-E_{Nn})O_{1N'm,Nn} \sim
\delta O_1 \delta U .
\label{uouo}
\end{equation}
If $N \ne N'$, so different energy shells separated 
by order $k \Delta U$ with $k > 1$, then
\begin{equation}
|E_{N'm} - E_{Nn}| > k \Delta U,
\end{equation}
which when combined with Eq. (\ref{uouo}), implies
\begin{equation}
\frac{O_{1N'm,Nn}}{\Delta O_1} < \frac{\delta O_1 \delta U}
{k \Delta U \Delta O_1} < \frac{\delta O_1 \delta U}
{\Delta U \Delta O_1} \ll 1 .
\label{o1do1}
\end{equation}
This shows that all off-diagonal matrix elements of ${\hat O_1}$ 
outside shell N are tiny, with those in
increasingly distant energy shells getting
increasingly smaller.
So to  a good approximation, all states outside the shell
N can be ignored and amongst the states, $\{\psi_{Nn}\}$,
in a given shell N, a new orthonormal basis 
$\{\phi^{(N)}_{N_1n}\}$ can be obtained by a unitary transformation
\begin{equation}
\phi^N_{N_1n} = \sum_{p \in N} \Psi_{Np} \langle \Psi_{Np}|\phi^{(N)}_{N_1n}
\rangle .
\label{pnnn}
\end{equation}
In the basis $\{\phi^N_{N_1n}\}$ and in a given cell $NN_1$,
${\hat H}$ is approximately diagonal with eigenvalue $U_N$ 
and ${\hat O_1}$
is approximately diagonal with eigenvalue $O^N_{1N_1}$.

In particular, for the new basis $\{\phi^N_{N_1}\}$,
for ${\hat H}$ it still holds 
\begin{equation}
\langle \phi^{N'}_{N'_1n'}|{\hat H}|\phi^N_{N_1n}\rangle
=[U^N + O(\Delta U)] \delta_{N'N} \delta_{N'_1N_1}\delta_{n'n}
+ O(\ll \Delta U),
\end{equation}
and now for ${\hat O_1}$ 
\begin{equation}
\langle \phi^N_{N_1n}|{\hat O_1}|\phi^{N'}_{N'_1n'}\rangle
= [O^N_{1N_1}+ O(\Delta O_1)]\delta_{NN'}\delta_{N_1N'_1} \delta_{nn'}
+ O(\ll \Delta O_1),
\end{equation}
where the magnitude of the off-diagonal elements of ${\hat O}_1$ are
determined as for the case of energy from
Eqs. (\ref{qmuncert}) and (\ref{dodo}) to
be for any $m$, $\sqrt{\sum_{n,n \ne m} O_{mn}^2} \ll \Delta O_1$.
In this new basis Eq. (\ref{pnnn}), with respect to the 
operators ${\hat H}$ and ${\hat O_1}$,
all states $\phi^N_{N_1n}$ in a given cell $NN_1$ are similar.
Finally, note that in the $\{\phi^N_{N_1n}\}$ basis, ${\hat H}$ also
has an ordering on its off-diagonal elements
from conditions analogous to 
Eqs. (\ref{uouo}) -(\ref{o1do1}).  In particular 
for sufficiently distant eigenstates 
(i.e. sufficiently large $k > 1$)
\begin{equation}
|O_{1m} - O_{1n}| \approx k \Delta O_1 ,
\end{equation}
from Eq. (\ref{commr})
\begin{equation}
(HO_1 - O_1H)_{mn} \approx (O_{1_m}-O_{1n})H_{mn} \sim
\delta O_1 \delta U ,
\end{equation}
which implies
\begin{equation}
\frac{H_{mn}}{\Delta U} < \frac{\delta O_1 \delta U}
{k \Delta O_1 \Delta U} < \frac{\delta O_1 \delta U}
{\Delta O_1 \Delta U} \ll 1 .
\end{equation}

Another observable ${\hat O_2}$ can now be considered.  Due to 
Eq. (\ref{dodo}) ${\hat H}$, ${\hat O_1}$, and ${\hat O_2}$ can be 
approximately mutually diagonalized.  In particular
the energy relation Eq. (\ref{commr}) applies equally to any operator
${\hat O_j}$ with the same consequence as found
in Eqs. (\ref{uouo}) - (\ref{o1do1}) 
for ${\hat O_1}$.  Next consider matrix elements
of ${\hat O_2}$ between two different $O_1$-cells
$NN_1$, and $NN'_1$. The same treatment applies here
as leading to Eq. (\ref{o1do1}), to give
\begin{equation}
\frac{O^N_{2N_1m;N'_1n}}{\Delta O_2} < \frac{\delta O_1 \delta O_2}
{\Delta O_1 \Delta O_2} \ll 1 ,
\end{equation}
when $N_1 \ne N'_1$.  Thus the basis states
$\{\phi^N_{N_1n}\}$ can be transformed to a new orthonormal
basis in which ${\hat O_2}$ is approximately diagonal in cells $N_2$
between which expectation values are separated by at
least $O(\Delta O_2)$, 
\begin{equation}
\phi^N_{N_1N_2m} = \sum_{p \in NN_1} \phi^N_{N_1p} 
\langle \phi^N_{N_1p}|\phi^{(N)}_{N_1N_2m} \rangle ,
\end{equation}
such that
\begin{equation}
\langle \phi^N_{N_1N_2n}|{\hat O_2}|\phi^{N'}_{N'_1N'_2n'}\rangle
= [O^N_{2N_1N_2} +O(\Delta O_2)] \delta_{NN'}\delta_{N_1N'_1}
\delta_{N_2N'_2}\delta_{nn'} + O(\ll \Delta O_2).
\end{equation}

This procedure of approximate diagonalization can be implemented
for all the operator observables under consideration
${\hat O_3}, {\hat O_4}, \ldots, {\hat O_j}$.
At the end, each energy shell $N$ is subdivided into cells
$J$ with the original approximate energy eigenfunctions
$\{\psi_{Nn}\}$ transformed to a new orthonormal basis
$\{\phi^N_{Jn}\}$ such that
\begin{equation}
\langle \phi^N_{Jn}|{\hat O_j}|\phi^{N'}_{J'n'}\rangle
= [O^N_{jJ}+ O(\Delta O_j)] \delta_{NN'} \delta_{JJ'} \delta_{nn'}
+O(\ll \Delta O_j).
\end{equation}
If thought of as matrices, the operators ${\hat O_j}$
in this basis are block diagonal, with each block
corresponding to a $J$-cell and with small nonzero matrix
elements connecting different blocks. 

\subsubsection{large number limit}
\label{largenl}

Consider a sequence of state vectors $\{\Psi_j\}$ 
in Eq. (\ref{psieca}) for which the
number density $N(x)$ in some region $x \in V$ 
Eq. (\ref{pnumber}) increases, and in turn for an
appropriate choice of basis states $\{ |\alpha \rangle \}$
the number of states in each cell increases. 
Suppose the matrix element                
\begin{equation}
\langle \Psi_j | {\hat  O} | \Psi_j \rangle,
\label{psiopsi}
\end{equation}
is computed for this sequence,  
where ${\hat O}$ is any one of the observables in the contracted description.
If there is no a priori
correlation amongst the states $|\alpha \rangle$ of the state
vector, as the number of components becomes large, cross terms in 
Eq. (\ref{psiopsi}) will cancel due to the random
phases of the expansion coefficients $c_{\alpha}$, whereas
the diagonal terms will not cancel.   Thus
\begin{eqnarray}
\langle \Psi_j | O | \Psi_j \rangle & = & \int \sum_{\alpha \beta}
O_{\alpha \beta} c_{\alpha} c_{\beta} \\
& \rightarrow & \int \sum_{\alpha} O_{\alpha \alpha} |c_{\alpha}|^2,
\end{eqnarray}
where $O_{\alpha \beta} \equiv \langle \alpha | O | \beta \rangle$.

This property is interpreted as a general outcome of the Correspondence 
Principle.  The Correspondence Principle states
that for a fixed measurement resolution, as the quantum numbers
of the observed system increase, the classical limit is reached.
One example of its application is the hydrogen atom.
which has energy levels $E_n \propto 1/n^2$ with the difference between
adjacent levels $\Delta E_n \propto (n+1)^{-2} - n^{-2}$.
The classical limit is when $\Delta E_n \propto 1/n^3$, which occurs
for large n when the leading quantum corrections
$\Delta E^{qn}_n \propto n^{-4}$ becomes negligible. 
For a given measuring apparatus that is
measuring the highly excited hydrogen atom,
as $n$ increases, eventually the resolution of the
apparatus can no longer discern the $n^{-4}$ quantum correction.
At that point,  one would
say the classical limit of the hydrogen atom has been reached up to the
measurement resolution of that apparatus.  In a many-body system with no
induced mechanism for attaining coherence, as the energy increases, it
implies either particles are being added to the system and/or the
particles in the system are being further excited.  In both cases, the
quantum numbers that are getting large are the particle occupancies of the
excited energy levels.  A measuring apparatus that probes the energetic
particles in the system will become increasing less sensitive to quantum
effects amongst the energetic particles as its detectors of fixed resolution
are forced to receive a
greater infux of particles.
This is the essence of the classical limit for a large many-body system.
In Sect. \ref{sect5} it will be shown that in this limit, quantum
evolution can be expressed in a form that resembles classical
evolution. 

\subsubsection{Transient state basis}
\label{transons}

The previous subsections have developed the concept of contracted 
descriptions and shown explicitly how to obtain them.
This subsection suggests both an appropriate Hilbert space
basis for the large transient state problem of heavy ion collision
as well as a contracted description within this basis.
In particular, following on the quark-gluon Fock space
basis of the quark-gluon plasma hypothesis, here we consider
this idea in its generalization as simply a Fock space  
of quasi-free particles which exist
during the lifetime of the transient state.
No assumption is required here that the underlying fundamental
theory, here QCD, is weakly interaction.
For strongly interacting dense systems it is possible that
more complicated excitations can emerge, which amongst themselves
interact with moderate or weak strength.
This quasiparticle picture \cite{migdal,anderson}
is successfully applied
to strongly interacting many-body electron systems.
In the present context, if a plasma like state is to emerge
during the course of a heavy ion collision, than
almost by definition
such a hypothesis implies that there
are particle-like excitations, whether quarks and gluons or
something else, existing during the lifetime
of the plasma phase, which do not interact so strongly that
they lose their individual particle identities.
Thus implicit in a plasma hypothesis is also
the assumption that some sort of quasiparticle excitations
emerge for some time period just after the collision. 
Of course in addition to these quasiparticle excitations
which define the properties of the plasma, in general
there would be other strongly interacting phenomenon.
To encorporate this basic picture,
the large transient state Hilbert space basis
$\{ \xi \}$ is constructed as a tensor product of transient states
$\{ \tau \}$ and core states $\{ c \}$,
\begin{equation}
\{ \xi \} = \{ \tau \} \times \{ c \}.
\end{equation}
Any state $| \tau\rangle$ in $\{ \tau \}$ by definition is
orthogonal to any state $|c\rangle$ in $\{ c \}$, 
\begin{equation}
\langle c | \tau \rangle =0,  \ \ \forall  \ \
|\tau \rangle \in \{\tau\}, \ |c \rangle \in \{c\}.
\end{equation}
The subspace $\{ \tau \}$ is a Fock space built from the set of
creation/destruction operators 
$\{ \tau^{\dagger}_{\alpha}({\bf k}, \tau_{\alpha}({\bf k})\}$,
where ${\bf k}$ is the momentum and $\alpha$ are
the internal quantum numbers of the transons.  
Additional restrictions may also be placed on
this transon Fock space.  For example, momentum regions
in which the transons are strongly interacting, and so not
well represented by the quasiparticle approximation
could be eliminated, such as the low momentum region in 
confining theories.  In \cite{te}, a separation of the Hilbert
space is also made similar to here, into what they
call the ``parton'' and ``gluonic'' subspaces, which have close
analogy to our transon and core subspaces respectively.

The reference state for the transon Fock space is defined by the
condition
\begin{equation}
\tau_{\alpha}({\bf k}) | {\rm ref} \rangle =0, \ \ \forall \ \
\tau_{\alpha}({\bf k}) \in \{\tau_{\alpha} \},
\end{equation}
and the states 
\begin{equation}
\tau_{\alpha}^{\dagger}({\bf k}) | {\rm ref} \rangle 
\end{equation}
define the excitations of single nearly-free particles with internal
quantum numbers $\alpha$, three-momentum ${\bf k}$, and 
energy $E_{\alpha}({\bf k}) = \sqrt{{\bf k}^2 + m_{\alpha}^2}$.
These nearly-free particles that comprise the subspace $\{ \tau \}$ will
be called transons.  The core states $| c \rangle$ are composed of all
states necessary to complete the transient state basis $\{ \xi \}$.
Note that by definition, any state in $\{ c \}$ can serve as the
reference state $| {\rm ref} \rangle$ for the transon Fock-space.
Our definition of the transon basis is not specific to transons
being weakly interacting, although as they are particle like
excitations, by definition their interactions are
moderate enough to preserve this property.
The idea here is similar to that for the parton subspace in \cite{te},
except in that work a more extreme condition is
imposed that interactions between the parton excitations
are completely switched off.

The contracted description of Subsect. \ref{contract} 
in the transient state
basis is as follows.  Recall the evolution of the transient state
wavefunction Eq. (\ref{psit}) is considered under the contracted description
Eq. (\ref{deloi}), and one seeks an appropriate set of
approximate eigenstates.  In the transient state basis, we
hypothesize the approximate eigenstates at energy $U$ up
to uncertainty $\Delta U$ are expressed as
\begin{equation}
\phi_{{\alpha}, \beta} (E_{\alpha} + E_{\beta}) 
\longleftrightarrow \prod_{j=1}^M 
\tau^{\dagger}_{\{\alpha_j \},\beta} ({\bf k}_j) 
| {\rm core_{\beta}} \rangle \ \ ({\rm general}) ,
\label{b2}
\end{equation}
where $\{ \alpha \} \equiv \{ \alpha_1 \ldots \alpha_M \}$
are the internal quantum numbers of the M-transons in this state,
$E_{\{\alpha \}} = \sum_{j=1}^M \sqrt{({\bf k}_j^2 + m_{\alpha_j}^2)}$,
and $E_{\beta}$ is the energy associated with the core state.
Note that the energy $E_{\alpha}$ associated with a state
$\phi_{\{\alpha\}, \beta}$ is only that of the transons.
This represents the energy from the state that is available for a large
many-body transient state.
It is also useful to identify two particular types of states in this basis

\begin{equation}
\prod_{j=1}^{M} \tau_{\alpha_j}^{\dagger}({\bf k}_j) | 0 \rangle_{\perp} 
\ \ \ \
\equiv | {\bf k}_{\alpha_1} \ldots {\bf k}_{\alpha_M} \rangle_{\perp}
\equiv |\{{\bf k}, \alpha \}^M \rangle \ \
({\rm pure \ \ transons})
\label{b1}
\end{equation}
and
\begin{equation}
| {\rm core} \rangle \ \ ({\rm pure \ \ core}). 
\label{b22}
\end{equation}
Here $|0 \rangle_{\perp}$ is the true vacuum of the theory, 
$|0 \rangle_{\rm true}$, except orthogonalized with respect to all pure
transon states
\begin{equation}
\label{vacperp}
|0\rangle_{\perp} = N \left[ |0 \rangle_{\rm true} -
\sum_{M=0}^{\infty} \sum_{\{\alpha\}}
\int \prod_{j=1}^M d^3k_j |\{{\bf k}, \alpha \}^M \rangle_{\perp} 
\langle \{{\bf k}, \alpha \}^M | 0 \rangle_{\rm true} \right],
\end{equation}
where $N$ is a normalization constant.
From the definition of similar states given below 
Eq. (\ref{psieca}), similarity
in the transon basis is defined to mean small separation
in all quantum numbers: occupancy of each
species, momentum distribution, and other internal quantum numbers
$\alpha$.  Depending on the type of measurement that is done, similarity
may also place conditions on the core states.

\section{Large Transient State Hypothesis of Heavy Ion Collision}
\label{sect4}

Based on the concepts of projection and contraction
developed in the last two
sections and combining that
with the implications of the large number limit, 
the large transient state hypothesis of heavy ion
collision now is stated.  Recall
in the time dependent
description, the two heavy ion wavepackets 
begin to overlap at $t \sim t_c^-$.
The nearly-free particles that comprise the transient state
must have the pseudo-fock space representation Eq. (\ref{b2}) in the collision
wavefunction $\Psi^{SE}(t)$ for $t \leq t_c^-$.
At the moment of collision,  the
components of $\Psi^{SE}(t_c^-)$ with
large occupancy of nearly-free particles are
associated in spacetime to the formation
of a dense many-particle region at the site of the collision.
The probability for creation of such a region is determined by the
probability amplitudes of the components of the wavefunction 
having high occupancy of
transons.  From Sec. \ref{largenl},
recall that for fixed measurement resolution,
the larger the state, the greater its tendency toward classical
behavior.
The large transient state hypothesis in the context of
heavy ion collision implies the existence of transon states
of adequately high occupancy, such that for measurement
resolutions of interest, their behavior is classical.
To classical characteristics of the large transient state
correspond properties of $\Psi^{SE}$: To entropy S corresponds
occupancy number, to temperature T corresponds
energy density and to energy U corresponds the sum of the free
particle energies of the nearly-free particles. Included
here is the spatial volume occupied by the transient
state, by combining T and U.

The wavefunction description of heavy ion collision,
relied upon heavily in this paper, clearly has limited
practical applicability in dynamical calculations due
to its complexity.  However, reference to wavefunctions
is advantages in developing the central theme of this paper,
which is to understand how classical behavior emerges in the
intrinsically quantum process of
heavy ion collision.  Furthermore, the use of wavefunctions
may be helpful in assessing probabilities and sizes of putative
large transient states, especially in relative terms,
such as p-p versus A-A or between ions of different nucleon
numbers.  

For example consider the following model.
Let the incoming p-p wavefunction just before the collision at $t_c^-$ 
be approximated by the product form
\begin{equation}
\Psi^{SE}_{pp} (t_c^-; ({\bf 0},{\bf k}), ({\bf 0},{\bf -k}))
= \Psi_{p^+}(t_c^-;{\bf 0},{\bf k}))  \Psi_{p^-}(t_c^-; ({\bf 0},{\bf -k}))
\end{equation}
where the expression $({\bf 0},k) \equiv ({\bf x},{\bf k})$ denotes
that the center of the proton p wavepacket at $t=0$ is at 
${\bf x} = {\bf 0}$ and the peak momentum along the
collision axis is {\bf k}.  The superscripts $+$,$-$ denote the direction of
the protons along the collision axis.
Each wavefunction has the quasi-Fock space expansion in
terms of transon and core states of the form
\begin{equation}
\psi_p(t; {\bf x},{\bf k}) = 
\sum_{M,\{\alpha\}} \sum_j
\int_{\infty} [d{\bf k}]^M
c_j(\{ {\bf k}, \alpha \}^M; t) |\{{\bf k}, \alpha \}^M, core_j \rangle,
\label{protonwf}
\end{equation}
where the shorthand notation used here and below is
\begin{equation}
\int_{\infty} [d{\bf k}]^M \equiv
\int_{-\infty}^{\infty} \prod_{j=1}^{M}
d^3{\bf k}_{j}
\label{dksh}
\end{equation}
and
\begin{equation}
\sum_{M,\{\alpha\}} \equiv
\sum_{M=0}^{\infty} \sum_{\alpha_1 \ldots \alpha_M} .
\label{sumsh}
\end{equation}
Thus for the heavy ion collision state, approximate it as the product state
of A + -moving and A $-$ -moving nucleons with 
bound state effects within each ion
represented only through assigning $A_c$ nucleons as central and $A_p$
nucleons as peripheral with $A=A_c+A_p$.  Thus the 
heavy ion A-A collision
wavefunction is approximated as
\begin{equation}
\Psi^{SE}_{AA} (t_c^-; ({\bf 0},Aq), ({\bf 0},-Aq))
= \Psi_{A_p}^+(t_c^-;({\bf 0},A_p q))  \Psi_{A_c}^+(t_c^-; ({\bf 0},A_c q))
 \Psi_{A_p}^-(t_c^-;({\bf 0},A_p q))  \Psi_{A_c}^-(t_c^-; ({\bf 0},A_c q))
\end{equation}
with 
\begin{equation}
\Psi^a_{A_l} (t_c^-; ({\bf 0},q)) =
\prod_{j=1}^{A_l} \Psi_{p_j}^a (t_c^-; ({\bf x_j},q)),
\end{equation}
$a = \pm$, $l=p,c$ and ${\bf x_j}$ the position of nucleon $p_j$ at 
time $t=0$.

Assume that a central collision is necessary to create a
large transient state in the central region
when the two heavy ions collide. Thus in what
follows the heavy ion collision is assumed to involve 
$\Psi^+_{A_c} \Psi^-_{A_c}$ with 
$\Psi^+_{A_p} \Psi^-_{A_p}$ passive.  
$A_c$ here is treated as a phenomenological parameter.

This model can demonstrate tendency
toward the large number
limit, thus based on Subsect. \ref{largenl}
the classical limit,
as $A$ increases.
For example, consider transient state formation in the central region. The
first step in the time history of such a process requires
a multiple collision event in which several quanta are stripped-off
both incoming heavy ions and deposited into the central
region.  The combinatoric probability for a multiple collision
event of this sort increases with $A_c$.
Suppose the initial collision between the two heavy ions
involves some type of quanta, such as quarks and gluons,
partons, hadronic resonances etc...  The precise nature
of these quanta does not need to be specified for this
analysis.  Simply suppose that $p(\sqrt{s})$ is the probability 
that two such quanta at CM energy $\sqrt{s}$ collide and deposit
all their energy into the central region.
Then the probability $P_{2m}(n)$
that out of $m$ impinging quanta from both directions
a collision with n-pairs occurs in which the energy is
deposited in the the
central regions would be
\begin{equation}
P_{2m}(n) = N_m(p) \left(\frac{m!}{(m-n)!n!}\right)^2 n!p^n ,
\label{probnm}
\end{equation}
where $N_m(p)$ is a normalization constant such that
$\sum_{n=0}^{m} P_{2m}(n) = 1$.
As an example, suppose the colliding quanta under consideration
were nucleons.  Let $p(\sqrt{s})$ be the probability in a 
nucleon-nucleon collision at CM energy $\sqrt{s}$
that energy deposition in the central region is above some
pre-specified density.  Then in a A-A collision, $A_c$ nucleons
in each impinging nuclei can in a central collision deposit their
energy in the same region, and the probability for this
would be from Eq. (\ref{probnm}),
$P_{2A_c}(A_c) = N_{A_c}(p) A_c! p^{A_c}$.
Thus for $p \ll 1$,
events with the maximum deposition of energy,
and so the largest transient states, would have
a highly suppressed probability, in fact
by several orders of magnitude, to the typical events.
For example at RHIC there are somewhere around 5 billion events recorded
in total and so for Au-Au collisions (A$=197$)
based on the above formula for $p \stackrel{<}{\sim} 0.9$
there would have been less than one high collision event
where all the 197 nucleons collided. 
This is a crude estimate but it demonstrates how suppressed
the very high collision events are, yet
these event fluctuations
may be the most interesting
for studying properties of large transient states.

\section{Derivation of the Master Equation}
\label{sect5}

This section obtains the evolution equation for the large transient state.
This equation will express the evolution of a single pure
quantum mechanical state up to an accuracy adequate for
measurement at the specified macroscopic resolutions 
$\{\Delta O_j\}$, Eq. (\ref{psieca}).
The large transient state by definition has a high density
of particles Eq. (\ref{pnumber}), so that within a resolution band
there are many quantum mechanical states.  Due to this, it will
emerge that the evolution equation for a pure
quantum mechanical state, representing a typical large transient
state, in general will have a classical form, in particular
the master equation.

Our derivation of this master equation follows Van Kampen \cite{vk}
(see Appendix A for a review),
except that we are converting his reasoning,
which addressed condensed matter systems, to
the scattering problem.
There are several differences between these two types of systems.
First, in the scattering problem the large transient state
exists only for a finite duration of time, whereas for the condensed
matter systems considered by Van Kampen \cite{vk},
this time interval effectively was infinite.
Second, in the scattering problem, the transient state of interest
is not in isolation.  Thus the entire scattering system can not be
subject to the master equation.  Third, the elementary excitations
of condensed matter systems generally are well understood,
since either they are closely associated with asymptotic particles
such as electrons or with lattice vibrations such as phonons.
In contrast in the scattering problem, since the elementary
constituents of QCD, quarks and gluons, confine, the elementary
excitations are unclear.  

The purpose of Sects. \ref{sect2} and \ref{sect3} was to clarify 
all three of these problems.
Points one and two were addressed through the
time-energy considerations in Subsect. \ref{timeup} and the definition
of projection operations in Subsect. \ref{proj}.
To address point three the problem was divided into two parts.
First the quasiparticles were defined in very general terms
via the transons of Subsect. \ref{transons}.
As a second step, specific identification of transons
with QCD excitations can be made.  This will be addressed
in Sect. \ref{sect6}.

The derivation that follows considers the simplest case in which the state
vector has an expansion in the pure transon
basis Eq. (\ref{b1}), with transons of scalar, spinor or vector types. 
It would
be straightforward to extend this derivation to
include the mixed basis of transon and core states.
The derivation below is done in a
conventional ${\bf k}$-space basis for the transons.  The same derivation
applies with minor notational changes if the kinematic specification of
the transons is in a mixed $({\bf x}, {\bf k})$-basis 
(see Appendix B) or in a basis of rapidity and transverse
momenta.
Turning to the derivation, let
\begin{equation}
c(\{{\bf k}, \alpha \}^M; t) \equiv
c({\bf k}_1,{\alpha_1}; \ldots ; {\bf k}_M,{\alpha_M}; t)
\end{equation}
denote the probability amplitude for the state with
$M$ transons of types
$\alpha_1 , \ldots, \alpha_M$ and respective momentum for $j$-th
transon,
${\bf k}_{j}$
\footnote{The momentum representation used here assumes the
probability to find two identical particles with exactly the same momentum
has zero measure.  For highly degenerate systems, like a Bose condensate,
an occupancy number basis is preferable to this momentum basis.
Our derivation to follow could be converted to an occupancy number
basis.}.
A transon type is characterized by all
internal quantum numbers: baryon number, spin, polarization, isospin,
and for nonasymptotic transons color, flavor, etc....  
The normalization  condition for the expansion coefficients is
\begin{equation}
\sum_{M,\{\alpha\}}
\int_{\infty} [d{\bf k}]^M
|c(\{ {\bf k}, \alpha \}^M; t) |^2 = 1 ,
\label{norm}
\end{equation}
where the shorthand notation is 
defined in Eqs. (\ref{dksh}) and (\ref{sumsh}).

A contracted description of the dynamics will now be constructed.
To illustrate this procedure, we hypothesize that the
the similar states of this contracted description
will be those states within small intervals of momentum to each
other and with some partitioning of the discrete quantum numbers.
The contraction over momenta is equivalent to
a course graining.  In particular, this will require dividing the
momentum interval into discrete cells with widths
$\Delta {\bf k} \equiv (\Delta k, \Delta k, \Delta k)$ in the three
spatial directions with the cells centered at points 
${\bf k}^{\bf m} \equiv ( {m_x}, {m_y}, {m_z}) \Delta k$
for integers $(m_x,m_y,m_z)$ and with each integer ranging from 
$-\infty$ to $\infty$.  In addition, the evolution will be examined in
discrete time intervals $\Delta t \ll \Delta t_{tr}$,
where $\Delta t_{tr}$ is defined in Subsect. \ref{timeup}
as the lifetime of the transient state. 
Accordingly the
probability coefficients will be smeared over a time interval $\Delta t$.

Denote the course grained probability coefficients as
\begin{equation}
P(\{{\bf k^{\{\bf m\}} }, {\bar \alpha}\}^M;t) \equiv
P({\bf k^{\bf m_1}},{\bar \alpha_1}; \ldots , 
{\bf k^{\bf m_M}}, {\bar \alpha_M}; t).
\end{equation}
Explicitly
\begin{equation}
\label{probdiv}
P(\{{\bf k^{\bf m} }, \bar \alpha \}^M;t) =
\frac{1}{\Delta t} \int_{t-\Delta t}^t dt'
\sum_{\{\alpha\} \in \{{\bar \alpha}\}}
\int_{\Delta^{\{\bf m\}}_{M}} [d {\bf k}]^M
|c(\{{\bf k}, {\alpha}\}^M,t')|^2 ,
\end{equation}
where
\begin{equation}
\int_{\Delta^{\{\bf m\}}_{M}} [d{\bf k}]^M\equiv
\prod_{j=1}^M 
\int_{k_x^{m_x^{j}}-\frac{\Delta k}{2}}^
{k_x^{m_x^{j}}+\frac{\Delta k}{2}} dk_{jx}
\int_{k_y^{m_y^{j}}-\frac{\Delta k}{2}}^
{k_y^{m_y^{j}}+\frac{\Delta k}{2}} dk_{jy}
\int_{k_z^{m_z^{j}}-\frac{\Delta k}{2}}^
{k_x^{m_z^{j}}+\frac{\Delta k}{2}} dk_{jz}.
\end{equation}
Note that
\begin{equation}
\sum_M \sum_{\{{\bf m}\}} \sum_{\{\bar \alpha\}}
P(\{{\bf k}^{\bf m}, \bar \alpha\}^M;t) =
\sum_{M=0}^{\infty}\sum_{\{\alpha\}} \int_{\infty} [d{\bf k}]^M
|c(\{ {\bf k}, \alpha \}^M;t) |^2 = 1.
\end{equation}

From the fundamental dynamics,  
we seek the evolution equation for the probability 
coefficients $\{P(\{{\bf k^{\bf m}}, {\bar \alpha} \}^M; t)\}$.
Furthermore we want to know their evolution between small finite time
steps $\Delta t \ll \Delta t_{tr}$ and not for infinitesimal time differences.
Let the initial state of the system, at say the time 
in Subsect. \ref{projdes}
designated to be just before the collision $t=t_c^-$,
be given as $\{c(\{{\bf k}, \alpha \}^M;t_c^-)\}$.  After a time 
$t > t_c^-$, the amplitude
coefficients of
the system will be
\begin{equation}
c(\{{\bf k}, \alpha \}^M; t) = \sum _{M', \{\alpha'\}}
\int_{\infty} [d{\bf k'}]^{M'}
\int \! \! \! \! \! \! \! \sum_r 
\langle \{{\bf k}, \alpha \}^M | r \rangle e^{-iE_r(t-t_c^-)} 
\langle r | \{{\bf k^{\prime}}, \alpha' \}^{M'}\rangle 
c(\{{\bf k^{\prime}}, \alpha' \}^{M'}; t_c^-) ,
\label{mserep}
\end{equation}
where
\begin{equation}
\int \! \! \! \! \! \! \! \sum_r |r \rangle\langle r| = 1 
\end{equation}
is an intermediate sum in the energy representation,
with this notation including the possibility of both discrete and
continuous spectra.  Note that this is the exact quantum evolution
of the system and for example no perturbative assumption 
of weak coupling is used.

From this we obtain the evolution of the course grained coefficients
to be
\begin{eqnarray}
P(\{{\bf k}^{\bf m}, \bar \alpha \}^M; t)&&=
\sum_{M' \{\alpha'\}} \sum_{M'',\{\alpha''\}} 
\sum_{ \{\alpha\} \in \{{\bar \alpha}\}}
\int_{\Delta^{\{\bf m\}}_{M}} [d{\bf k}]^M
\int_{\infty} [d{\bf k}^{\prime}]^{M'}
[d{\bf k}^{\prime\prime}]^{M''}
\nonumber \\
& & \left[\{{\bf k}, \alpha \}^M || \{{\bf k}^{\prime}, \alpha^{\prime}\}^{M'};
\{{\bf k}^{\prime\prime}, \alpha^{\prime \prime}\}^{M''}\right]_{t-t_c^-}
c(\{{\bf k}^{\prime}, \alpha^{\prime}\}^{M'}; t_c^-)
c^{*}(\{{\bf k}^{\prime\prime}, \alpha^{\prime \prime}\}^{M''};t_c^-) ,
\label{pevol}
\end{eqnarray}
where
\begin{eqnarray}
\left[\{{\bf k}, \alpha \}^M || \{{\bf k}^{\prime}, \alpha^{\prime}\}^{M'};
\{{\bf k}^{\prime\prime}, \alpha^{\prime \prime}\}^{M''}\right]_t
& \equiv & {1\over\Delta t} \int_t^{t+\Delta t}\! \! \! \! dt^{\prime} 
\int \! \! \! \! \! \! \! \sum_{r}
\int \! \! \! \! \! \! \! \sum_{s}
\langle \{{\bf k}, \alpha\}^M|r\rangle 
\langle r|\{{\bf k^{\prime}}, \alpha^{\prime}\}^{M'} \rangle
\nonumber \\
& & \langle \{{\bf k}, \alpha \}^M|s\rangle^* 
\langle s|\{{k^{\prime\prime}}, \alpha^{\prime \prime}\}^{M''} \rangle^*
\exp{[-i(E_r-E_s)t]}.
\label{mematrixe}
\end{eqnarray}
%and,
%\int_{\Delta^m_n} [dk]_n \equiv 
%\int_{k_1^{m_1}-{\Delta_1\over2}}^{k_1^{m_1}+{\Delta_1\over2}} dk_1 \ 
%\ldots \ \int_{k_n^{m_n}-{\Delta_n\over2}}^{k_n^{m_n}+{\Delta_n\over2}} dk_n  
%\eqno(3-425)
%$$

Similar to the arguments in Subsect. \ref{largenl},
in the limit that the system size is large, there will be
many cross terms in Eq. (\ref{pevol})
and the relative phases amongst the amplitude coefficients will
be random initially and thus during evolution.  Therefore the cross terms
will tend to cancel.  On the other hand,
the diagonal terms are all positive,
and thus will always add.  This is the first element of
Van Kampen's theory.
From this we get
\begin{eqnarray}
P(\{{\bf k}^{\bf m}, \bar \alpha\}^M; t) = 
\sum _{M^{\prime},\{\alpha^{\prime}\}}
\sum_{\{\alpha\} \in \{{\bar \alpha}\}}
\int_{\Delta^{\{\bf m\}}_{M}} [d{\bf k}]^M
\int_{\infty} [d{\bf k}^{\prime}]^{M^{\prime}}  & &
\left[ \{\bf k, \alpha\}^{M} |
| \{{\bf k^{\prime}}, \alpha^{\prime} \}^{M'};
\{\bf k^{\prime }, \alpha^{\prime}\}^{M'} \right]_{t-t_c^-}
\nonumber \\ & &
\left| c(\{\bf k^{\prime}, \alpha^{\prime} \}^{M'};
(t_c^-)\right| ^2 .
\end{eqnarray}

The second observation of Van Kampen
was that within a course
grained cell $\Delta^{\bf m}_M$, both factors in Eq. (\ref{pevol})
vary independently
\footnote{The point can be more clearly illustrated with a discrete sum.
Consider the sets of N positive numbers $\{a_j\}$ and $\{b_j\}$.
Let $a \equiv {{\sum a_j} \over N}$ and
$b \equiv {{\sum b_j} \over N}$.  Define $\epsilon_j^a$ and $\epsilon_j^b$
such that $a_j=a+\epsilon_j^a$ and $b_j= b+\epsilon_j^b$.  This implies
by definition that $\sum \epsilon_j^a= \sum \epsilon_j^b = 0$.
Then we have $\sum a_j b_j=
\sum (a+\epsilon_j^a)(b+\epsilon_j^b)=
Nab+\sum \epsilon_j^a \epsilon_j^b \rightarrow
Nab$.
The last step follows since the second term involving the sum of
$\epsilon_j^a \epsilon_j^b$ will have both positive and negative
contributions.  When N is large these will tend to cancel
amongst each other.}.
This implies
\begin{eqnarray}
P(\{k^{\bf m}, \bar \alpha\}^M; t) &=&
\sum_{M^{\prime},\{\alpha^{\prime}\}} \left(
\sum_{\{\alpha\} \in \{\bar \alpha\}}
\int_{\Delta^{\{\bf m\}}_{M}} [d{\bf k}]^M
\int_{\Delta^{\bf m'}_{M^{\prime}}}
\left[{d{\bf k}^{\prime}\over\Delta}\right]^{M'}
\left[\{{\bf k}, \alpha \}^M ||\{{\bf k^{\prime}}, \alpha^{\prime}\}^{M'};
\{{\bf k^{\prime}}, \alpha^{\prime}\}^{M'}\right]_{t-t_c^-} \right)
\nonumber \\
& & \left(\int_{\Delta^{\bf m'}_{M'}}
\left[d{\bf k}^{\prime}\right]^{M'}  |
|c(\{{\bf k}^{\prime}, \alpha^{\prime}\}^{M'}; t_c^-)|^2\right) 
\nonumber \\
& = & \sum_{M'\{\bar \alpha'\}} \sum_{\{\bf m'\}}
[\{{\bf k^{\bf m}}, \bar \alpha \}^M || \{{\bf k^{\bf m'}}, \bar \alpha' \}^{M'}
\{{\bf k^{\bf m'}}, \bar \alpha' \}^{M'}]_{t-t_c}
P(\{{\bf k^{\bf m'}}, \bar \alpha'\}^{M'},t_c),
\label{psep}
\end{eqnarray}
where
\begin{equation}
\int_{\Delta^{\{\bf m\}}_{M}} 
[\frac{d{\bf k}}{\Delta}]^M\equiv
\prod_{j=1}^M 
\int_{k_x^{m_x^{j}}-\frac{\Delta k}{2}}^
{k_x^{m_x^{j}}+\frac{\Delta k}{2}} \frac{dk_{jx} }{\Delta k}
\int_{k_y^{m_y^{j}}-\frac{\Delta k}{2}}^
{k_y^{m_y^{j}}+\frac{\Delta k}{2}} \frac{dk_{jy}}{\Delta k}
\int_{k_z^{m_z^{j}}-\frac{\Delta k}{2}}^
{k_x^{m_z^{j}}+\frac{\Delta k}{2}} \frac{dk_{jz}}{\Delta k}
\end{equation}
and
\begin{equation}
[\{{\bf k^{\bf m}}, {\bar \alpha}\}^M || \{{\bf k^{\bf m'}}, {\bar \alpha'}\}^{M'}
\{{\bf k^{\bf m'}}, {\bar \alpha'} \}^{M'}]_{t} \equiv
\sum_{\{\alpha\} \in \{\bar \alpha\}}
\int_{\Delta^{\{\bf m\}}_{M}} [d{\bf k}]^M
\int_{\Delta^{\bf m'}_{M^{\prime}}}
\left[\frac{d{\bf k^{\prime}}}{\Delta}\right]^{M'}
\left[\{{\bf k}, \alpha \}^M ||\{{\bf k^{\prime}}, \alpha^{\prime}\}^{M'};
\{{\bf k^{\prime}}, \alpha^{\prime}\}^{M'}\right]_{t} .
\end{equation}
Note the above is a statement about the dynamics and not
the initial conditions.

We can now obtain the evolution equation for the course grained
probabilities.  For this we are interested in time steps
$\Delta t$ and not for infinitesimal time differences. 
For small $\Delta t$ upon
temporal averaging and course graining one obtains the general form
\begin{eqnarray*}
[\{{\bf k^{\bf m}},\bar \alpha \}^M || \{{\bf k^{\bf m'}}, \bar \alpha'\}^{M'}
\{{\bf k^{\bf m'}}, \bar \alpha' \}^{M'}]_{\Delta t} =
\delta_{M M^{\prime}} \delta_{\{\bar \alpha\} \{\bar \alpha'\}}
\delta_{ \{\bf m\}\{\bf m'\}}
+ \Delta t W\left(\{{\bf k}^{\bf m}, \bar \alpha \}^M;
\{{\bf k^{\bf m^{\prime}}}, {\bar \alpha'}\}^{M'} \right) .
\label{smallt}
\end{eqnarray*}
For convenience, converting the notation now to a single index,
\begin{equation}
(M,\{\bar \alpha\},\{\bf m\}) \Longrightarrow J
\end{equation}
the evolution equation for the probability coefficients becomes
\begin{eqnarray}
\frac{dP_{J}(t)}{dt} & \equiv &
\frac{P_J (t+\Delta t) -
P_{J}(t)}{\Delta t} 
= \sum_{J^{\prime}}
W_{JJ^{\prime}}
P_{J^{\prime}}(t) .
\label{peom}
\end{eqnarray}
Above, we have used Eq. (\ref{psep}) 
to express $P_J(t+\Delta t)$
in terms of the set $\{P_J(t)\}$.  To arrive at
the above form, one should also note that 
$P_{J'}(t)$ is expressed
in the form of Eq. (\ref{probdiv}).
In Eq. (\ref{peom}) $W_{JJ'}$ denotes
the transition probability between states J and J' and
$P_J(t)$ denotes the occupation probability of state J.

Since
\begin{eqnarray*}
\sum_{M} \sum_{\{\bf m\}} \sum_{\{\alpha\}}
[\{{\bf k^{\bf m}},\bar \alpha \}^M || \{{\bf k^{\bf m'}}, \bar \alpha'\}^{M'}
\{{\bf k^{\bf m'}}, \bar \alpha' \}^{M'}]_{t} = 1 ,
\end{eqnarray*}
%\eqno(3-10)
this implies 
$\sum_J W_{JJ'}=0$.
%\eqno(3-11)
From this we arrive at
\begin{equation}
{dP_J(t) \over {dt}} = \sum_{J'} \left[W_{JJ'}P_{J'}(t)-W_{J'J}P_J(t)\right],
\label{mastereqn}
\end{equation}
%\eqno(3-12)
which is the master equation.
%This equation is derived under the condition that there is no
%coherence between the amplitude coefficients in the transon
%basis.

Let us comment on some aspects of Eq. (\ref{mastereqn}).
The equilibrium distribution is determined by the solution of
\begin{equation}
\sum_{J'} \left[(W_{JJ'}P^e_{J'}(t)-W_{J'J}P^e_J(t)\right] =0 .
\end{equation}
%\eqno(3-13)
Also in equilibrium, by the equipartition hypothesis it follows
that,
\begin{equation}
\frac{P^e_J(t)}{P^e_{J'}(t)} = \frac{G_J}{G_{J'}},
\label{equimult}
\end{equation}
%eqno (3-13.25)
where $G_J$ is the total multiplicity of the states in the 
cell $J$.
The proof of detailed balance follows from time reversal
symmetry of the underlying dynamics.
Let $|{\rm T}(\{{\bf k},\alpha\})\rangle$ be the time reversed state of
$|\{{\bf k},\alpha\}\rangle$. Then it follows from microscopic dynamics that,
\begin{equation}
\langle \{{\bf k'}, \alpha'\}; t | \{{\bf k}, \alpha\}, 0 \rangle=
\langle T(\{ {\bf k}, \alpha \} ); t | T(\{ {\bf k'}, \alpha' \}), 0 \rangle .
\end{equation}
%(3-14)
Since a course grained cell always contains pairs of state related by
T, it means that
\begin{equation}
W_{JJ^{\prime}} G_{J^{\prime}} =
W_{J^{\prime}J} G_{J} .
\label{timerel}
\end{equation}
%(3-15)
Eqs. (\ref{equimult}) and (\ref{timerel}) give the necessary ingredients to
obtain the detailed balance relation
\begin{equation}
W_{JJ'}P^e_{J'}=W_{J'J}P^e_J ,
\end{equation}
%eqno(3-16)
where there is no summation on the repeated indices.

Let us now address when the canonical thermal distribution can be applied
to describe the transient state.  Recall that the canonical distribution
describes a system in equilibrium when it is in contact with an
external heat bath.  In contrast, the transient state is an isolated
system at some fixed energy.  An ensemble respecting such
an isolated system in equilibrium consists then of all
states $r$ within a small interval $\Delta U$
about the given energy $U_r$ having equal probability,
%\begin{equation}
%P_r = \left\{ \begin{array}{cl}
%             C & \mbox{if $E_r - \delta E < E < E_r + \delta E$}\\
%             0      & \mbox{otherwise}
%             \end{array}
%     \right.
%
%P_r = C \ {\rm if} \ E_r - \delta E < E < E_r + \delta E , \ O \ otherwise,
%\label{mcd}
%\end{equation}

\begin{equation}
  \label{mcd}
  P_r = \left\{
\begin{array}{ll}
             C & \mbox{if $U_r - \Delta U < U < U_r + \Delta U$} \\
             0      & \mbox{otherwise} .
\end{array}
\right.
\end{equation}
The canonical distribution can be applied to such a microcanonical
ensemble Eq. (\ref{mcd}) for describing some small part of the whole
system.  For example suppose we separate the cell $J$ into a small
portion denoted by $J_S$ and the remainder $J_R$,
$J_S \bigcup J_R = J$ and $J_S \bigcap J_R = 0$.
We can obtain probability coefficients exclusively
for $J_S$ by summing over all configurations of $J_R$,
so $P_{J_S} = \sum_{J_R} P_J$.
The cells $J_S$ are no longer closed systems at some fixed energy,
since energy can flow from $R$ into $S$.
Then provided $R$ is very large compared to $S$,
by the usual system-reservoir arguments \cite{reif,kittel},
in equilibrium the $P_S$ coefficients will satisfy
the canonical distribution.
For example if our cells were in x-space, we could ask what is
the equilibrium distribution of some small volume of
the complete transient state, and in equilibrium it would
obey the canonical distribution.
Aside from the equilibrium distribution, the master equation
permits very general characterization of the statistical state;
an example of some other possibilities which may be interesting
to consider in the context of the heavy ion transient states
are \cite{mestates}.
An examination of how pure quantum evolution in systems with
a large number of degrees of freedom develops into a statistical
mechanical behavior has been studied for general systems
in \cite{srednicki}.  There are similarities between
that work and ours here, such as similar conditions are
considered on diagonal versus off-diagonal matrix elements.
Moreover our conclusions are similar to \cite{srednicki}
although we obtain a general evolution
equation, the master equation, that can treat a variety
of statistical behavior whereas in \cite{srednicki} the focus
is only on understanding the equilibrium limit.

This Section examined the evolution equation of a pure
quantum state which is measured with macroscopic uncertainties,
meaning resolution widths in which a large number
of quantum states are indistinguishable.  For such a case,
it was shown above that the evolution equation 
for the state vector has a classical
form, in particular the master equation.
This formal association with the master equation implies all
properties of this equation in its application
to nonequilibrium statistical systems carry over.
In particular, recall the master equation 
keeps account of all transitions between the states of
a statistical system, and thus is
the fundamental equation
for describing the nonequilibrium dynamics of the given system.
Moreover, other contracted descriptions, further down in
the hierarchy can be obtained from it, such as the
Boltzmann equation and the hydrodynamic equation \cite{vkbook}.
As one detail, rather than the k-space basis for the transon
excitations, as used in this Section, another possibility
is a basis of phase space excitations, as described in
Appendix B, for which  the resulting
master equation would express transitions in phase
space.  Such an equation formally would be analogous
to that for a classical gas, for which considerable
information about master equation solutions already exist
and could be readily utilized.
Alternatively the transons could be described in terms
of rapidity in the longitudinal direction and
two transverse momenta. Such a choice of kinematic variables
would be a more appropriate starting point for deriving
the Bjorken hydrodynamical model \cite{bj} from a more
fundamental level using the master equation
transcription of this paper.

In closing this Section, an important distinction is emphasized
here between pure quantum states versus statistical ensembles.
A pure quantum state contains information about one single scattering
event.   In contrast a statistical ensemble contains information
over some type of summation over several events, each
represented by a pure state.  The Wigner distribution \cite{wigner,carzac}
is an approach to examine evolution equations of
statistical ensembles of quantum mechanical pure states. 
However, the origin of the master equation derived
here has no relation to the Wigner distribution.  The
justification for classical arguments, such as quark-gluon
plasma formation, in predicting outcomes of a scattering
experiment, must be established on individual
events and that is what our derivation of the
master equation has accomplished.

\bigskip
\section{Heavy Ion Collision Scenario}
\label{sect6}
\medskip

This section reviews the phenomenology 
of transient state formation during heavy ion collision
and then explores how it
could be studied
using the methods developed in this paper.
Although the heavy ion collision scenario was mainly
formulated with the quark-gluon plasma formation scenario in mind,
after our review in Subsect \ref{phic}, the picture is reexamined in the more
general terms developed in the previous sections.

\subsection{Phenomenology of heavy ion collision}
\label{phic}

The basic scenario for large transient state formation
in heavy ion collision was formulated in
\cite{bj,mcl}, in terms of a spacetime picture and this will now
be reviewed.
The process of interest is $A-A$ collisions at ultra-relativistic
energy.
The events of specific interest for exhibiting statistical 
mechanical behavior are central collisions; such collisions can be
identified by an enhanced production of grey-prongs \cite{greyp}.
Alternatively at RHIC central collisions are characterized
by a combination of both minimum number of particles in
the zero-degree calorimeters (ZDC) and 
the largest event multiplicities \cite{centralex}.
In spacetime, central collisions are described to occur at impact
parameter b less than the range of the nuclear force $b < 1 {\rm fm}$.
In the center-of-momentum frame, both relativistic nuclei appear
Lorentz contracted along their longitudinal direction of motion,
so that at extremely high energies, the collision is
likened to that of two discs or ``pancake-like'' 
projectiles \cite{car1,bj}.
The spacetime picture provides a center-of-momentum per nucleon 
energy scale, implicit to the Lorentz contraction of 
$E_{CM}^0 \approx 7 - 70 {\rm GeV}$, above which the longitudinal
contraction is smaller than 
the characteristic hadron scale $\sim 0.1-1 {\rm fm}$.
Phenomenologically at center-of-momentum energies above  $E_{CM}^0$,
it is expected that the collision will result in two
fragmentation regions approximately separated from a 
central region.  

The heavy ion experiments of interest for
quark-gluon plasma formation are at large nucleon
number $A \stackrel{>}{\sim} 100$ and at center-of-momentum energies 
above $E_{CM}^0$ per nucleon, since such cases are
believed to yield
the highest probability for
realizing a large energy density central region.
Heavy ion facilities meeting these specifications were initially
the CERN SPS collider where $\sqrt{s} \approx 20{\rm GeV}$ per nucleon,
at present is RHIC 
where $\sqrt{s} = 200 {\rm GeV}$/nucleon and in the 
near future will be ALICE
where $\sqrt{s} = 5.5 {\rm TeV}$/nucleon.
The spacetime evolution picture in this energy
regime is as follows.  The colliding nuclei are
rather transparent to each other. However the low momentum components
will interact strongly and some portion may come to rest 
at $t \approx t_c^+$ in the
central region about $z \approx 0$.  
This scraped-off debris will continue to interact with
constituents of high longitudinal momentum in the receding nuclei
after $t > t_c^+$.  These interactions will occur after a time
$\tau \sim 1 {\rm fm/c}$ in the rest frame of the colliding 
constituents.  This implies the higher the longitudinal momentum,
the later they interact and the further away from the collision
point $z \approx 0$ they are deposited.  
One phenomenological fact from pp-collisions that is
carried over for A-A collisions is the presence of particle
distribution in the central region that is uniform in rapidity.
Combining this with the above spacetime picture,
it implies new matter of increasingly larger rapidity is deposited
at the edges of the central region, just behind the receding
nuclei, and when looked within the rest frame of this matter,
the energy density is constant.
In addition, the two separating nuclei may "heat-up" as they
pass through each other and this could lead to baryon rich fireballs in
the two fragmentation regions \cite{akm} (but see also \cite{lead}).  

The above picture supplies the initial conditions on the matter
in the central region.  To treat the subsequent evolution, another 
phenomenological input from pp-collisions is applied, that
there is little long range correlation in rapidity.
This transcribes in spatial terms to mean interactions amongst
particles produced in the central region only extending
a small distance $\Delta z \approx O(1) {\rm fm}$ along the
collision axis.  Thus each slice of matter in
the central region of thickness
$\Delta z \approx O(1) {\rm fm}$ evolves primarily independently
except for some interactions with the nearest neighbor slices.
Since matter first is deposited at $z \approx 0$, the centermost
slice begins its evolution first, with slices increasingly further
away from $z =0$ beginning evolution at increasingly later
time.  Initially the particles deposited in the central
region have a large distribution of longitudinal momentum,
whereas the transverse momentum distribution is random.
This leads to a longitudinally expanding central
region with initially no transverse expansion.
The quark-gluon formation scenario is focused
on the intermediate time period just after the central and
two receding nuclei have
approximately separated from each other, 
at $t \geq t_c^+$ and before 
final state hadronic particle
production has occurred at $t < t_h$.
The time when the central region and two receding nuclei
separate into three independent regions
is estimated as $t_c^+ \approx 1fm/c$, where we have set the
initial time of impact as $t_c^- \approx -1fm/c$.

The evolution of the
central region, in particular, is modelled as an isentropic hydrodynamic
expansion \cite{bj,kap}. 
In this picture, the phenomenologically motivated assumption 
based on $pp$ collisions is that the particle distribution
in the central region just after impact, say at $t=t_c^+$, is
uniformly distributed in rapidity
sets the initial conditions for the subsequent evolution
of the central region. In particular, at $t=t_c^+$, the particles in
the central region, which corresponds to 
a small region at say around ${\bf x}=0$, have a large dispersion
in velocities, with the longitudinal velocity gradient being large, whereas
the transverse velocity dispersion is random.

Estimates \cite{bj,mcl} of the central region place its
volume as $V \approx (5-100) fm^3$
and energy density as $\epsilon \approx (1-10) {\rm GeV}/{\rm fm^3}$.
This leads to a total energy in the central region of
$U \approx (5-1000) {\rm GeV}$.  The central region is
pictured to expand for a duration of time in a plasma-like
state, before hadronization begins at $t_h$.
Estimates give $t_h/t_c^+ \approx 10-1000$, which
implies the plasma state extends for a time interval
of $\Delta t_{tr} \approx (10-1000)fm/c$.
Due to the high energy density of the plasma, the collision
time for its particles is estimated as 
$\tau_{collision} \sim (0.1-1) fm/c$, whereas the expansion
time scale of the central region, since it is effectively
a 1-d expansion, is $V/{\dot V} \sim t \approx 1fm/c$.
From comparing these two scales, it is assumed the local
dynamics within the plasma are only weakly
affected by the system's expansion.

This scenario now will be interpreted in terms of the formalism
developed in the previous sections.  
The projection step of Subsect. \ref{proj}
is accomplished here by going to sufficiently high collisions
energies, $\sqrt{s} > E_{CM}^0$, where the final state particle
production in the central region are well separated from those in the
two fragmentation regions.  The large transient states of interest 
are believed to form for a selected set of events in the
central region, which are distinguished through additional
conditions placed on the final state particle spectrum.  To specify these
conditions, the contraction step is necessary.
For this the full set of operator observables and
measurement uncertainties must be specified and then,
as discussed in Subsect. \ref{timeup}, theoretical models of detail and
accuracy consistent with experimental resolution must be formulated.
Sect. \ref{sect3} explained the purpose and methodology of the contraction
step and finished-off in Subsect. \ref{transons}
by introducing the transon model, which is a generalized
basis and contracted description for the heavy ion problem.
Finally Sect. \ref{sect6} determined the evolution
equation for a quantum state vector expressed in terms of the
transon basis.  Specifically, relying on the large number limit,
it was demonstrated there how the quantum evolution equation
reduces in form to the master equation of nonequilibrium
statistical mechanics.  The initial conditions for evolving
the transient state must be constrained by the specifications
of the above reviewed phenomenological model.
With this, the master equation then can dynamically determine
how the state evolves, whether towards thermalization
or some other statistical state.
 
\subsection{Identifying transon candidates}
\label{transonp}
 
What remains and will now be discussed is the specifications
of the excitations that comprise the transons.
In broader terms, 
if a large manybody transient state
formed during a heavy ion collision, it would rapidly evolve through several
stages in which several types of excitations may exist. The appropriate
question is does any particular manybody excitation dominate during the
transient period sufficiently to leave some imprint on the final state?
The central step in answering this question is to determine the types of
QCD excitations that can be the transons of a large transient state during a
heavy ion collision. Some of the possibilities are as follows:

{\it Transons as hadrons} - This is the benchmark example
which form the constituents of the hadron gas transient 
state \cite{hagedorn,carlitz,Rafelski:1984mq} . Since hadrons
and hadronic resonances are asymptotic states, their properties
are unambiguous.  Therefore, statistical mechanics is applicable to
treating the hadron gas with few assumptions.

{\it Transons as partons} - 
The parton model is used to treat quark-gluon interactions
in perturbative QCD high energy 
collision problems, where
both incoming hadrons exchange only a few of these  particles.
Thus the parton model is applicable for short lived, small transient states
with very high energy densities.  On the other hand, in the large
transient state heavy ion collision problem, both incoming
heavy ions exchange many constituent particles.  This problem
addresses moderately high energy densities distributed over
a large spatial volume.  The only similarity  between both
problems is that individual collisions between the quarks
and gluons is at adequately high energies for asymptotic freedom
to be applicable \cite{gm}.   Aside from this, both problems differ
substantially.   Although probabilistic distributions appear
in pQCD problems, their  interpretation is completely different
from those associated with the master equation of Sect. \ref{sect5}.
In pQCD, the probabilistic distribution functions for
the quarks and gluons arise from an averaging over
a certain class of events, which exploits unitarity to
eliminate infrared divergences.  In contrast,  in the large transient
state problem, only one single event is treated, and the cancellation
of any infrared divergences must arise from within that
single system.

{\it Transons as energetic ($|{\bf p}| \ge T$
quarks and gluons} - This appears to be the democratic
hopeful. The picture is that of a high temperature gas of
quarks and gluons, which due to asymptotic freedom will interact weakly.
Two questions are generally raised about quark-gluon plasma.  The first
question pertains to the existence of such a phase in QCD. The second
question is whether proper conditions can be realized during a heavy ion
collision, to produce such a phase.  The first question has been
studied extensively by both lattice simulations \cite{qcdptlat}
and various analytical approaches \cite{qcdptanal}. The evidence suggests
that QCD undergoes a phase transition from a gas of hadrons to a gas of
quarks and gluons at a temperature $T_c = 190 {\rm MeV}$. The properties
of the quark-gluon plasma have been examined 
considerably \cite{qgp,experiment,mcl,bj,propqgp},
with particular emphasis on a gauge invariant description.
The second question is much less clearly understood, since
there are a few complicating factors to this simple picture. First,
a perturbative treatment of such a state leads to infrared
divergences \cite{linde}, which generically are expected in any theory with
a massless gauge boson.  Optimistically this is a technical problem of
infrared summation.  Considerable effort has gone to study 
this \cite{irsolve}.
Second, a transient state of energetic quarks and gluons that
is produced in the laboratory is immersed within the true 
vacuum of QCD $|0 \rangle_{\rm true}$.
However perturbative high temperature QCD calculations on the
thermodynamics of quarks and gluons are performed with respect to the
perturbative vacuum of QCD $|0 \rangle_{\rm pert.}$.  
Is this theoretical idealization
well representative of the actual process? 
in the transon model, this question equates to whether
$| 0 \rangle_{\rm true}$ and $ | 0 \rangle_{\perp}$ are "approximately" the
same, where 
$|0\rangle_{\perp}$ from Eq. (\ref{vacperp}) is 
the true vacuum orthogonalized with
respective all energetic quark-gluon states that primarily comprise the
large transient state.  If affirmative, it means the true
vacuum does not contain many energetic quarks and gluons and so
$|0\rangle_{\perp}$ should be approximately stationary. 
The transon model can also be approximated by replacing
$0\rangle_{\perp}$ with $|0\rangle_{\rm pert.}$.  This replacement is valid
if the energetic quarks and gluons of the transient state interact
minimally with the low energy components of the true vacuum.  High
temperature perturbative QCD in conjunction with the 
bag model \cite{mitbag}
follows this set of assumptions with the additional requirement
that there is a global shift in energy 
$E_{\rm pert.} - E_{\rm true} = B$.
The final dilemma in interpreting quarks and gluons as the transons is that
the picture of a color singlet transient state represented in a Fock space
of quarks and gluons is misleading since the gluon field operator
$A^{a\mu}(x)$ does not transform in color space as an octet.  At the level of
thermodynamics this does not appear to be a big problem,
since the differences are not significant between a
color unrestricted 
versus color-singlet projected summation of the partition
function \cite{kapcol},
yet out-of-equilibrium this could be a more important issue.

{\it Transons as manybody quasiparticle} - The essential question
in realizing quark-gluon plasma is whether color can flow over extended
regions.  If that occurs, it implies a liberation of color degrees of
freedom, which is the signature of the new thermodynamic state.  For this
to occur, it is not mandatory that the gas of nearly-free particles be
quarks and gluons.  The effective dynamics that most conveniently
characterizes color flow may require collective variables of the entire
system as opposed to the fundamental quark-gluon operators.  In analogy
to condensed matter physics systems, these collective variables
generically can be called quasiparticles. 

In condensed matter systems a plasma oscillation is a collective
longitudinal excitation of the elementary particles and one quantum of
this excitation is called a plasmon.  Finite temperature perturbative QCD
calculations indicate a similar collective 
effect amongst the gluons \cite{brpi}. 
The calculation of the finite temperature gluon self-energy show a nonvanishing
longitudinal and transverse gluon mass. This modifies the gluon
dispersion at low momentum from $E=|{\bf p}|$ to
$E=\sqrt{{\bf p}^2 + m_g^2 + O({\bf p}^2)}$. Here $m_g \sim {\rm MeV}$ 
is the induced
gluon mass or synonymously the plasmon frequency.
A color singlet collective density fluctuation that propogates through the
plasma at sound velocity has also been considered in
\cite{phonon} and identified as the hydrodynamic phonon.

{\it Transons as elementary color singlet clusters} -
We can regard hadrons and liberated quarks and gluons as two
extreme ends of the range of possible excitations in the large
transient state.  It is possible that the for the large
transient state created in heavy ion collision, a mixture
of both color singlet and color liberated excitations may
co-exist.  In statistical systems, it is common that the long range order
in the system is in a certain phase, but there is short range order
with properties of an entirely different phase.  For example,
for the Ising model in the high-temperature phase, there is no
long range order, since globally the spins are decorrelated.
Nevertheless, locally correlations amongst nearby spins
still exist, with this short range order being increasingly
more pronounced as the critical point is approached from above.
In analogy, the large transient states produced in heavy ion collision
may not have adequately high energy density and a sufficiently
large volume to totally eliminate local color correlation.
In this case color singlet clusters will emerge.  Their density
may be quite low, so that just a few small color singlet clusters are
immersed in an otherwise sea of liberated quarks and gluons.
On the other hand, the whole phase may be dominated by color singlet
regions of size bigger than hadron scales but smaller than the volume 
of the transient state.

The importance of neutralizing color in the large transient state appears
from another direction, which further motivates the presence
of color singlet clusters.  Due to confinement in QCD,
all energy eigenstates of the theory are color singlet.
Since evolution of the scattering process 
could be described in Eq. (\ref{mserep})
with an energy eigenstate basis,  it means in principle
the system can be described with only color singlet excitations.
Moreover, assuming an adequate separation 
between the central and fragmentation
regions, the transient state in the central region, described by
the projection Eq. (\ref{pnumber}) would be color singlet, and so by similar
reasoning, its excitations also could be described with only
color singlet states.  An immediate consequence of these
considerations is the transient state basis in Subsect. \ref{transons}
should be restricted to color singlet states.  There
are a few ways in which this could be realized.  One
possibility is the transon and core states independently
are both color singlets.  A second possibility is the transons,
most likely high energy quarks and gluons, be regarded in
a color unrestricted Fock space, where the core states
appropriately neutralize color.
The interpretation is that an arbitrary color unrestricted state
of energetic quarks and gluons will neutralize itself through low energy
components, which are represented through the core states.  If it is then
assumed that the low energy components have no other affect on the
development of the large transient state, then they can be dropped.
With further assumptions,
The color unrestricted quarks-gluon states can be constructed over a
convenient reference state such as $|0 \rangle_{\rm pert.}$.
Finally, a third possibility is the individual transon excitations
themselves are color singlet.  If the transon excitations
were hadrons, that would be one example of this.  However,
it is possible that more elementary color singlet clusters of just a few
correlated quarks and gluons,
could emerge over the time scale of the transient state.
A color singlet quark-gluon Hilbert state also can be constructed,
%by use of
%the gluon field tensor 
%$F^{a \mu \nu}(x) \equiv \partial^{\mu} A^{a \nu}(x) -
%\partial^{\nu} A^{a \mu}(x) -ig\epsilon^{abc}A^{b\mu}(x) A^{c\nu}(x)$,
%which transforms as a color octet.  In this case, 
in which color singlet composite
operators are in general products of quarks and gluons
\begin{eqnarray}
&  & O^{\mu_1 \ldots \mu_n}_{f_1 \ldots f_M {\bar f}_1 \ldots {\bar f}_N}
({\bf R})  =  \int d^3r_1 \ldots d^3r_{M+N} 
\delta^{(3)}({\bf R} - \sum_{j=1}^{M+N} \frac{{\bf r}_j}{N+M})
F({\bf r}_1 \ldots {\bf r}_{M+N})
\nonumber \\
& & \prod_{j=1}^M \psi^{f_j}_{a_j v_j}({\bf r}_j)
\prod_{j=1}^N {\bar \psi}^{{\bar f}_j}_{b_j w_j}({\bf r}_{j+M})
\Gamma^{\mu_1 \ldots \mu_n}_{v_1 \ldots v_M, w_1 \ldots w_N}
\prod_{i=1 \ldots M,j=1 \ldots N} U^{a_i, b_j}({\bf r}_i,{\bf r}_{j+M}) ,
\end{eqnarray}
%\begin{eqnarray}
%O_{LMN}^{ww'}({\bf R}) \equiv
%\int & & \prod_{l=1}^{L+M+N-1} d^3{\bf r}_{ll+1}
%f_{s}({\bf r}_{12},{\bf r}_{23},\ldots,{\bf r_{m+n-1m+n}},{\bf R})
%\nonumber \\
%& & \prod_{k=1}^{L} F^{a_k \mu_k \nu_k}({\bf r_k}
%\prod_{i=1}^M {\bar \psi}^{C{\bar a}_i{\bar \delta_i}}_{{\bar f}_i}
%({\bf r}_{L+i})
%\prod_{j=1}^N {\psi}^{C a_j \delta_j}_{f_j}({\bf r}_{L+M+j})
%\Gamma_{cw}^{o_1, \ldots, a_M}
%\Gamma_{rw'}^
%{\mu_1\nu_1, \ldots, \mu_L\nu_L,{\bar f}_1, 
%\ldots, f_n,{\bar \delta}_1,\ldots,\delta_n}
%\label{qcomp}
%\end{eqnarray}
%where ${\bf r}_{ll+1} \equiv {\bf r}_{l+1}-{\bf r}_l$
%and ${\bf R} = (\sum_{i=1}^{m+n} {\bf r}_i)/(m+n)$,
where $\Gamma_{cw}^{\mu_1 \ldots \mu_n}$ is a Dirac tensor
associated with the $M$ fermions of flavor $f_j$ and $N$ anti-fermions
of flavor ${\bar f}_j$
and $U^{ab}({\bf r}, {\bf r'})$ are the gauge field link matrices.
With these operators, the product state
\begin{equation}
\prod_{j=1}^{n} O({\bf R}_j) |) \rangle_{\perp}
\label{qgcomp}
\end{equation}
are not orthogonal, although if their centers ${\bf R}_j$
are sufficiently far apart, they should be approximately orthogonal.

\subsection{Examples of models}

In order to do a concrete calculation with the methodology developed
in this paper, what is needed first is transon and core
states to be identified and a Hamiltonian specified which
describes their interactions.  With this in place,
matrix elements, etc... required for the master equation
in Sect. \ref{sect5} can then be computed.  One also
would need to calculate the eigenstates and eigenfunctions for
this system, which in most practical cases would have to
be done perturbatively.  

The ultimate goal would be to start with the full QCD Hilbert space
of states and then identify the transon excitations that would
be relevant during the transient state created during the
heavy ion collision, with some possibilities listed
in Subsect. \ref{transonp}.  With this identification, the
full QCD Hilbert space is then decomposed into the transon and
core sectors and an effective Hamiltonian is obtained that
describes all interactions. With this done, one can proceed with
construction of the master equation following Sect. \ref{sect5}.

However this is too ambitious for the time being. A more immediate
application of the methods in this paper would be to provide
a more complete dynamical description of some of the hypothesized
models of the transient state such as the hadron gas etc...
For the case of the hadron gas, rather than just computing thermodynamic
quantities, a dynamical nonequilibrium treatment is possible.
For example, one could consider the situation where the
primary transon states are the $\pi$, $\rho$ and $a_1$ mesons.
These mesons are considered important in the production
of dileptons and photons in the hadron gas \cite{efflagdpprod}.
One could start with the effective Lagrangian for this
system \cite{efflagdpprod}, and compute in perturbation theory the
energy eigenstates, eigenvalues and matrix elements
needed in Eq. (\ref{mematrixe}) from which the master equation is
constructed.  Finally one needs to specify an initial state,
when these excitations become important in the
transient state, from
which to start evolution of the master equation.
This might be specified through considerations
of hydrodynamical models \cite{bj} and quark and gluon
parton distributions \cite{gm,qpic}.

In a similar way, the nonequilibrium dynamics of other
possible transon excitations can
be examined via the master equation.  For example, strange quark production
could be examined in chiral models \cite{efflagstrprod}.
Alternatively, effective theories of 
quarks and gluons \cite{efflagqg}
could be examined as simplified versions of
a complete QCD treatment of quark-gluon plasma.

\bigskip
\section{Conclusion}
\label{sect7}
\medskip

The purpose of this paper was to
surmise the first principles origin of a classical statistical
mechanical description of multi-hadron production in heavy ion
collision. Some special focus was placed on quark-gluon plasma formation,
due to its widespread interest.
As we have seen, realizing a statistical mechanical description
requires a straightforward application of fundamental and well known
concepts about the large number limit and randomness.  The master
equation, which is the final result of this transcription, lends
itself to a large variety of statistical mechanical behavior, of which
the often studied case of thermalization is just one special case.  

The transcription made in this paper was for a single pure quantum state
of two incoming heavy ions. This is not the familiar starting point for
treatments of heavy ion collision.
Normally a density matrix formalism is used \cite{carzac,elzeheinz}, due to
the belief that is the only practical means of treating such
a complex system.
However, in principle
one should be able to construct a pure quantum state of two incoming nuclei,
and studies its evolution
in a manner no different from say electron-electron scattering.
Furthermore if one is making predictions about final state observables
such as particle yield and type, they must have their explanation
at the level of pure quantum dynamics.  This has been one motivation
for the considerations in this paper.

This paper addressed a similar problem to that in \cite{te}, which is
under what conditions does a pure quantum state behaves in a classical
manner, with application to the heavy ion collision problem. Our
decomposition of the full Hilbert space into transons and core states is
analogous to the parton and gluonic subspaces respectively
identified in \cite{te},
with the diagonalization in the partonic subspace in \cite{te} similar but
more extreme to our approximate diagonalization in the transon subspace.
Where this paper goes beyond \cite{te} is in addressing
time evolution.  In \cite{te} the development stops at the point where the
decomposition into their parton and gluonic subspaces is made, in
which by construction observables in the partonic subspace are diagonal, i.e.
the Schmidt decomposition. As stated there, such a construction in general
can only be made at one time slice.  In order for their decomposition to
sustain over time the perspective adopted in \cite{te} was this is a model
specific issue depending on the particular
nature of the interactions in the fundamental Hamiltonian.
In particular it depends on whether in the particular theory
one is examining certain states, called "pointer" states, can be identified
which in essence would preserve the Schmidt decomposition over time. Thus
based on the perspective in \cite{te}, since it is dynamics specific, the
conclusion is that in general the Schmidt decomposition will not remain
valid over time.  The key step where this paper deviates
from \cite{te} is that nowhere in \cite{te} are the facts
exploited that
they are examining a system with a large number of degrees
of freedom and that ultimately a contracted description
of the system will be made.  What we observe in this paper is that in the
large number limit, a search for exact pointer states from the fundamental
Hamiltonian is not needed.  In the large number limit, the conditions on the
interactions can be relaxed considerably to only that they are not very
strong, and in such a case, the large number limit will dominate over
any off-diagonal correlations thus realizing an evolution that is
effectively classical. The main result in this paper was 
in Sect. \ref{contdy},
which demonstrated how this approximate diagonalization occurs in the large
number limit. In the language of \cite{te}, our results basically
say that there are a class of states that effectively can act as pointer states
and identifying this class
depends on the specified degree of resolution to which the system
is measured.  Thus in general the set of pointer states is not unique, but
rather a whole range of states can emerge which within the needed resolution
of the measurement can behave effectively as the pointer states.
It also should be noted here that although there are many
treatments of classical limits to quantum mechanics via
path integral approaches, our derivation in Sect. \ref{contdy}
as an extension and completion of that started by
Van Kampen in \cite{vk}, is one of the few that
demonstrates the classical limit within a completely
canonical approach; independent of
our application to the heavy ion collision problem,
this is another significant result of this paper.

Historically, thermodynamic approaches to hadron scattering in
the "old" school were initiated by Fermi \cite{fermi}
and developed also by Landau \cite{landau}.
In more recent times Carruthers and collaborators \cite{car2,car4} 
have developed these
ideas to a more general nonequilibrium statistical mechanical
approach.  In this "new" school of thought, the change
in perspective has been that there need not be anything less
rigorous about using statistical mechanical variables
as the fundamental modes of description \cite{car1}.
Although theorist have studied more seriously 
the predictive content of statistical kinetic theory to scattering
processes, at present little unification with 
conventional quantum mechanical approaches to scattering
has been made.  This paper has attempted to shed some light
in this direction.  For one thing, this paper provides a general
framework in terms of the threefold projection, contraction, evolution,
scheme, which converts a quantum mechanical
scattering problem involving a large composite system into a form
that resembles a nonequilibrium statistical mechanical problem.

Our purpose here was to convey an alternative mode of thinking, that 
ultimately encapsulates a more general methodology.  In doing this,
the treatment in this paper has only captured a small essence of the problem
and this after considerable simplification.
To briefly summarize, in our approach, first, time was separated
into three sections, initial, final, and interacting.
Ascribing an initial and final interaction time implicitly
relies on a description with wave-packets.
We then froze the initial state at the onset of the interacting time period.
Due to the largeness of the initial system, we assumed randomness
amongst the expansion coefficients when expressed in terms
of the quasiparticle basis and assumed this held
throughout evolution in the interacting
time period.  In the limit that this approximation is
valid, dynamics took a form of classical probabilistic evolution.
How valid is this limit in a real world problem?
There obviously is not complete randomness in the initial state, since
a great part of it propogates into the final state.  How should one
describe the initial state as an expansion of weakly interacting
quasiparticles plus a coherent core?
An approach was suggested in Subsect. \ref{transons} in terms of 
what we called core states and generalized particle like excitations 
termed transons.
How universal is such a separation for different heavy nuclei?

Although the answers to such questions will be hard to answer, these
questions are demonstrative of a rich variety of physics made available
by a statistical mechanical description of multi-hadron production.
In as much as high-energy hadron scattering is a manybody problem,
the high energy collision of two heavy ions is a many-manybody problem.
The path via statistical mechanics toward its understanding could be
summarized as an attempt to replace s,t, and u with S,T, and U.

\bigskip

\section{Acknowledgements}

\medskip 

%Dedicated in the memory of Peter Carruthers.
I thank Peter Carruthers for many helpful discussions during
`93 and `94, when the ideas for this paper were developed.
I also thank the following for helpful discussions,
J. Collins, T. Lappi,
S. Mandelstam, J. Rafelski, D. Soper, M. Strikman, and D. Toussaint.
Financial support was partially provided by
the US Department of Energy and the UK Particle Physics
and Astronomy Research Council (PPARC).

\section*{Appendix A}
\label{appa}

In this appendix Van Kampen's
theory is reviewed.  Given a Hamiltonian, ${\hat H}$, 
Van Kampen first separates its
eigenstates into groups.
Each group is specified by a central energy $U_j$ and a resolution
$\Delta U_j$.  The resolution is fixed by the experimental apparatus.
He then classifies
suitable linear combinations of eigenstates within an energy band,
into equivalence classes.  To implement the classification, he
introduces the set of all operator
observables that the experimental apparatus measures.  This set
includes the Hamiltonian, ${\hat H}$, and
all other observed operators $\{{\hat O}_1,..., {\hat O}_j\}$.
The resolution of all observables 
$R=\{\Delta U;\Delta O_1, \Delta O_2, \cdots, \Delta O_j\}$ 
are set by the experimental apparatus.
All states within the resolution band R and with central
values $C_l\equiv\{U_l;O_{1l}, ...,O_{jl}\}$ are placed in
equivalence class $G_l$.  

To obtain the set of states in $G_l$, Van Kampen starts with the set
of eigenstates within the band $U_l \pm \Delta U$ and proceeds
constructively as follows.  He forms linear combinations from these
states, so as to diagonalize the operator ${\hat O}_1$ up to resolution
$\Delta O_1$.  Within each band $O_{1l} \pm \Delta O_1$,
he repeats his "partial diagonalization" procedure with
respect to the operator ${\hat O}_2$ and so on for the rest. In the end he
ends up with certain linear combinations amongst the
eigenstates in the specified resolution band.  These states are grouped
into the equivalence classes $G_l$.  The
states in each equivalence class $G_l$, up to observation,
are identical.  All systems measured with resolution R about
central values $C_l$ have a Hilbert space expansion 
solely with respect to the states
in $G_l$.

The energy operator played a special role in that the
system at all times remained within its initially specified energy
band.  The problem of dynamics to solve
here was, given an initial state with coarse grained expectation
values $\{O_{1l},\cdots,O_{nl}\}$, what are the expectation values
at a future time t.  The macroscopic nature of the problem enters because
the observables were specified within resolution bands
much bigger than the minimal values set by quantum dynamics.
Under these specifications, Van Kampen attempted to derive an evolution
equation.

It is helpful to state the above in a slightly different way also.
The initial state is first specified to lie is an
invariant  subspace of the complete
Hilbert space, based on the resolution widths of all
conserved observables.  Temporal evolution is therefore
restricted to lie on the so determined manifold of states.
The description of the state vector is, however, further described
with respect to a set of nonconserved observables.
It is the temporal evolution of these quantities that constitutes
the dynamical, nonstationary aspect of the problem.

In relativistic scattering processes,  in addition to energy, the conserved 
observables are the total three-momenta, the angular momentum
and discrete quantum numbers, such as baryon number etc...
The nonconserved "observables" in the construction of the main text
were the number density operators of the quasiparticle states.

Having specified the initial state within equivalence classes, $G_l$,
rather than immediately introduce a statistical density
matrix, Van Kampen followed a slightly different approach,
which retains closer contact to the underlying
quantum dynamics.  Van Kampen constructed a pure state
by taking an arbitrary linear combination of states in $G_l$
as the initial state.  Since the expansion coefficients have no
specified correlations amongst each other for the states in $G_l$,
in the limit the number of states, $N_l$ in $G_l$ gets bigger
he showed how dynamical evolution took the form of a master equation.
We followed this reasoning in our explicit construction
in Sect. \ref{sect5}.
With this slight chance in viewpoint, Van Kampen was able to
demonstrate a matter of principle which is often obscured in
treatments that start on the onset with a statistical density matrix.
In particular, he found a set of conditions under which
quantum dynamics goes over into a form that resembles
nonequilibrium statistical dynamics.

\section*{Appendix B}
\label{appb}

An alternative to the k-space transon excitations
used in Sect. \ref{sect5} is phase space
localized excitations.  These are defined as

\begin{equation}
\tau^{\dagger}_{\alpha}({\bf x}_r,{\bf p}_s)=\int d^3{\bf k} 
g_{\{{\bf x}_r,{\bf p}_s\}}({\bf k}) \tau^{\dagger}_{\alpha}({\bf k}) ,
\label{waveletop}
\end{equation}
where $\tau^{\dagger}_{\alpha}({\bf k})$ is a creation operator for a 
transon with three-momentum ${\bf k}$ and with all other quantum
numbers represented through $\alpha$.
%To construct quasiparticle creation operators
%$q^{\dagger}_j(x_r,p_s)$, we use,
%\begin{eqnarray*}
%M_{ij}(\bar{R})=\int d^3\bar{r}\ F_m(\bar{r})\left[\bar{\psi}_{ia}^\alpha
%\left(\bar{R}-{\bar{r}\over 2}\right)\ U_{ab}\left(x-y\right)\ 
%\Gamma_{\alpha\beta}\ \psi_{jb}^\beta \left(\bar{R}+{r\over 2}\right)\right]
%\end{eqnarray*}
%$$
%\eqno(6-1a)
%$$
%\begin{eqnarray*}
%G(\bar{R})=\int d^3\bar{r}\ F_G(\bar{r})\left[ G^a_{\alpha\beta}
%\left(\bar{R}-{\bar{r}\over 2}\right)\ G^a_{\delta\sigma}
%\left(\bar{R}+{\bar{r}\over 2}\right)\
%\Gamma^{\alpha\beta\delta\sigma}
%\right] \cr
%\end{eqnarray*}
%$$
%\eqno(6-1b)
%\begin{eqnarray*}
%B_{ijk}(\bar{R})&=&\int d^3r_1\ d^3r_2\ d^3\bar{r}_3\ F_B
%\left(\bar{r}_1,\bar{r}_2,\bar{r}_3\right)\left[\psi^{\alpha}_{ia}
%\left(\bar{R}-r_1\right)\ \psi^{\beta}_{jb}\left(\bar{R}-\bar{r}_2\right)
%\right.\cr
%&&\left. \psi^\gamma_{kc}\left(\bar{R}-\bar{r}_3\right) U_{am}(\bar{r}_1)\ 
%U_{bn}(\bar{r}_2)\ U_{co}(\bar{r}_3)\ \Sigma^{abc}\ \Gamma_{\alpha\beta\gamma}
%\right]\cr
%\end{eqnarray*}
%$$
%\eqno(6-1c)
%$$
The set  $\{g_{({\bf x}_r,{\bf p}_s)}(\bf k)\}$
is a complete set of phase space centered orthogonal functions
with centers $(x_r,p_s)$ defined on an integral lattice of spacing a and b
as $(x_r,p_s)=(ra,sb)$.
Some examples from wavelet theory would be the generalized Wannier
functions \cite{wilson} and the Gabor functions \cite{gabor,djj}.   
Note that the
phase space points form a discrete not continuum set,
with the volume of each cell governed by the uncertainty principle.

With the choice of transon states in Eq. (\ref{waveletop}), the continuum
integrals in Eq. (\ref{norm})
and the equations that follow are all replaced
by discrete sums over the phase space points $\{x_r,p_s\}$. 
The contraction procedure in this case would group
states in nearby phase space points into similar states,
along with additional conditions imposed by grouping
of the other quantum numbers $\alpha$.  Thus, when
the state vector expansion in this wavelet basis
satisfies the conditions for the master equation evolution, these enlarged
phase cells will correspond to classical statistical cells.
With the master equation expressing transitions amongst statistical
cells, the problem formally becomes equivalent to
that of a classical gas of particles.  For this case,
considerable information is available about treating master equations.
In particular,
following standard lines now from classical statistical mechanics, 
the Boltzmann equation can be derived
from Eq. (\ref{mastereqn}), using for example
Siegert's construction \cite{siegert}.  
Then from the Boltzmann equation, the 
hydrodynamic equations can be obtained by known methods \cite{vkbook}.

\end{document}